\documentclass[aps,prfluids,twocolumn,showkeys,showpacs,amsmath,amssymb,superscriptaddress, longbibliography, nofootinbib]{revtex4-2}
\usepackage[bookmarks=true, colorlinks=true, linkcolor=RoyalBlue, urlcolor=blue, citecolor=OliveGreen, pdftex, bookmarks=true, hyperindex=true]{hyperref}

\usepackage{graphicx}
\usepackage{url}

\usepackage{amsmath}
\usepackage{amssymb}

\usepackage{color}
\usepackage[dvipsnames]{xcolor}

\usepackage[percent]{overpic}
\usepackage{combelow}

\usepackage{lipsum}
\usepackage{xparse}

\newcommand{\cdv}[2]{\dfrac{\mathrm{D}#1}{\mathrm{D}#2}} %
\usepackage{physics}        %
\usepackage{tensor}         %
\usepackage{siunitx}        %
\usepackage{acro}           %
\usepackage{tikz,calc}      %
\usepackage{pgfplots}		%
\pgfplotsset{compat=newest}
\usepackage{tikz-3dplot}    %
\usepackage{cancel}         %

\definecolor{orange}{rgb}{1,0.5,0}
\definecolor{forestgreen}{rgb}{0.13, 0.55, 0.13}
\definecolor{bittersweet}{rgb}{1.0, 0.44, 0.37}
\definecolor{chartreuse}{rgb}{0.87, 1.0, 0.0}
\definecolor{darkorchid}{rgb}{0.6, 0.2, 0.8}

\DeclareAcronym{hp}{short = HP, long = Hermite polynomial}
\DeclareAcronym{nse}{short = NSE, long = Navier-Stokes equations}
\DeclareAcronym{lbm}{short = LBM, long = lattice Boltzmann method}
\DeclareAcronym{lbe}{short = LBE, long = lattice Boltzmann equation}
\DeclareAcronym{bgk}{short = BGK, long = Bhatnagar-Gross-Krook}
\DeclareAcronym{cfd}{short = CFD, long = computational fluid dynamics}
\DeclareAcronym{rk3}{short = RK3, long = third-order Runge-Kutta}
\DeclareAcronym{vlbm}{short = vLBM, long = vielbein lattice Boltzmann method}
\DeclareAcronym{weno5}{short = WENO-5, long = fifth-order weighted essentially non-oscillatory}

\begin{document}

\title{Vielbein Lattice Boltzmann approach for fluid flows on spherical surfaces}

\author{Victor E. Ambru\cb{s}}
\affiliation{Department of Physics, West University of Timi\cb{s}oara, Bd.~Vasile P\^arvan 4, Timi\cb{s}oara 300223, Romania}

\author{Elisa Bellantoni}
\affiliation{%
Computation-based Science and Technology Research Center, The Cyprus Institute,
2121 Nicosia, Cyprus
}
\affiliation{%
Department of Physics \& INFN, Tor Vergata University of Rome, 
00133 Rome, Italy
}%
\affiliation{%
LTCI, T\'{e}l\'{e}com Paris, IP Paris,
91120 Palaiseau, France
}%

\author{Sergiu Busuioc}
\affiliation{Department of Physics, West University of Timi\cb{s}oara, Bd.~Vasile P\^arvan 4, Timi\cb{s}oara 300223, Romania}

\author{Alessandro Gabbana}
\email[]{alessandro.gabbana@unife.it}
\affiliation{CCS-2 Computational Physics and Methods, Los Alamos National Laboratory, Los Alamos, 87545 New Mexico, USA}
\affiliation{Center for Nonlinear Studies (CNLS), Los Alamos National Laboratory, Los Alamos, 87545 New Mexico, USA}
\affiliation{Università di Ferrara and INFN-Ferrara, I-44122 Ferrara, Italy}

\author{Federico Toschi}
\affiliation{Department of Applied Physics and Science Education, Eindhoven University of Technology, 5600 MB Eindhoven, The Netherlands}
\affiliation{CNR-IAC, I-00185 Rome, Italy}

\begin{abstract}
    In this paper, we develop a lattice Boltzmann scheme based on the vielbein formalism 
    for the study of fluid flows on spherical surfaces. 
    The vielbein vector field encodes all details related to the geometry of 
    the underlying spherical surface, allowing the velocity space to be treated 
    as on the Cartesian space. The resulting Boltzmann equation exhibits inertial (geometric) forces, 
    that ensure that fluid particles follow paths that remain on the spherical manifold,
    which we compute by projection onto the space of Hermite polynomials. 

    Due to the point-dependent nature of the advection velocity in the polar coordinate $\theta$, 
    exact streaming is not feasible, and we instead employ finite-difference schemes. 
    We provide a detailed formulation of the lattice Boltzmann algorithm, 
    with particular attention to boundary conditions at the north and south poles.

    We validate our numerical implementation against two analytical solutions of the Navier-Stokes equations 
    derived in this work: the propagation of sound and shear waves. Additionally, we assess the robustness 
    of the scheme by simulating the compressible flow of an axisymmetric shockwave 
    and analyzing vortex dynamics on the spherical surface.
\end{abstract}

\maketitle

\section{Introduction}\label{sec:intro}

Fluid flows on curved manifolds are ubiquitous to several physical systems in nature,
ranging from interface rheology in foams~\cite{cox-ra-2004}, the dynamics of confined active matter~\cite{henkes-pre-2018, nitschke-prf-2019, rank-pof-2021}, geophysical flows of oceanic and atmospheric circulation~\cite{sukoriansky-prl-2002,schneider-ams-2009,deplace-sc-2017} up to exotic applications in the study of the motion of electrons in 2D materials~\cite{giordanelli-sr-2018} or quark-gluon plasma formed in heavy-ion collisions \cite{Romatschke:2011hm,Ambrus:2022adp}.

The spherical geometry holds particular relevance for geophysical flow modeling, 
as complex three-dimensional flows in atmospheres and oceans are typically 
modeled assuming planetary surfaces can be approximated as spheres. 
Global atmospheric circulation on Earth and large planets is commonly
modeled by two-dimensional Euler or Navier-Stokes equations (often in the 
vorticity-stream function formulation~\cite{nitschke-jfm-2012, reuther-mms-2015, gross-jcp-2018}) 
coupled to classic thermodynamics, quasi-geostrophic equations, and rotating shallow water equations~\cite{zeitlin-book-2018}. 
Notably, spherical geometry has been identified as the optimal testing ground 
for exploring two-dimensional turbulence through means of numerical simulations~\cite{lindborg-jfm-2022}.
These flows are characterized by large vortex structures due to the inverse energy cascade,
combined with scale filaments due to the enstrophy cascade, and it is well known that, under
real-world parameter ranges, the energy spectrum extends over several orders of magnitude, 
making a direct numerical simulation (DNS) approach computationally unfeasible. 
This calls for the development of new, efficient and accurate numerical methods leveraging
the ever-growing computational power available on modern massively parallel architectures.
In addition, analytical solutions for fluid flows on the sphere are rare~\cite{supekar-jfm-2020}, which makes it generally difficult to benchmark and compare the performances of various numerical schemes.

The contributions provided in this work are twofold: i) we define two benchmarks for axisymmetric flows
on the sphere, presenting exact analytic solutions of the Navier-Stokes equations, and ii) we
introduce a \ac{lbm} formulation based on differential geometry.

LBM has emerged as a highly efficient and flexible mesoscopic approach
for the simulation of many different fluid flows~\cite{succi-book-2018}.
Although its standard formulation is based on the use of Cartesian grids, 
in the past decades several formulations have been proposed to extend the 
applicability of the method to curvilinear coordinates.
One first approach~\cite{mendoza-sr-2013, budinski-cf-2014, yoshida-jcp-2014, debus-sr-2017}
consists in moving the metric dependence from the streaming part of the lattice Boltzmann equation
to the forcing and collision part, adding additional terms that can be identified by performing
a Chapman-Enskog expansion to match the target macroscopic equations. 
The advantage of this approach is that exact streaming is preserved; however, the resulting scheme typically supports only mildly curved surfaces.
An alternative approach consists of making the advection velocity space coordinate dependent.
No extra source terms are required. However, the price to pay is the loss of exact streaming,
hence requiring interpolation for the implementation of the advection step, which introduces
extra numerical dissipation. Most of these schemes are based on finite difference or
finite volume methods~\cite{fan-pesca-2005, dhiraj-jcp-2009, li-ccp-2016, yang-ec-2022}.
Recently, this approach has been employed in an elegant formulation based on differential
geometry and vielbein fields, closely related to the definition of kinetic theory 
in general relativity~\cite{cardall-prd-2013}, which allows for 
advection velocities to become coordinate independent~\cite{busuioc-pre-2019,busuioc-jfm-2020}, in principle simplifying the task of handling more complex geometries.

In this work, we adopt the vielbein approach to define an LBM for describing 
fluid flows on the spherical surface. We assume radial fluid motion 
is suppressed by internal or external centripetal forces 
(e.g., attraction between a spherical membrane and biological fluid above it), 
allowing the description of vector fields and their gradients using differential geometry.
The vielbein fields separate the local geometry from the definition of vector field components. 
This is convenient when describing the local phase space. 
The conceptual simplicity of the vielbein scheme makes it a promising candidate for tackling more complex geometries, 
such as those relevant to biomembrane modelling~\cite{baumgart-2003-nat}, including the case of time-dependent surfaces~\cite{torres-sanchez-2019-jfm, voigt-2019-jfm}.

For example, the square of the particle 
velocity shows an explicit geometry dependence when expressed with respect to the spherical 
coordinate basis, $\mathbf{v}^2 = (v^\theta)^2 + \sin^2\theta (v^\varphi)^2$. On the contrary,
with respect to the vielbein components $(v^{\hat{\theta}}, v^{\hat{\varphi}}) = (v^\theta, 
\sin\theta v^\varphi)$, the square of the velocity becomes coordinate independent:
$\mathbf{v}^2 = (v^{\hat{\theta}})^2 + (v^{\hat{\varphi}})^2$.
We validate our scheme through numerical benchmarks, examining the dynamics of sound
and shear waves while comparing the accuracy and convergence order with different interpolation
methods for the advection step. We further test our scheme in compressible flows using
a Sod shock tube-like setup that produces propagating and reflecting shock waves. 
We solve these axisymmetric problems using both axisymmetric and rotated grids, 
requiring fully 2D simulation for the latter, in order to validate our scheme's isotropy. 
We conclude by studying two-vortex dynamics on a spherical surface.

The remainder of this paper is organized as follows. In Section~\ref{sec:hydro}, we present the hydrodynamic equations for fluid flows on a sphere.
In Section~\ref{sec:method}, we present all the steps required to develop and implement a vielbein LBM solver specialized to fluid flows on the sphere.
Readers primarily interested in the numerical results may proceed directly to Section~\ref{sec:numerics}, where we present numerical benchmarks and assess the accuracy of the method. This is followed by a brief analysis of computational performance in Section~\ref{sec:performances}.
Finally, concluding remarks and future directions are summarized in Section~\ref{sec:conclusions}. 

For completeness and to support reproducibility, additional technical details are provided in the appendices.
Appendix~\ref{app:quad} introduces the $Q = 3$ and $Q = 4$ quadrature models, also known as the $D2Q9$ and $D2Q16$;
while both recover the continuity and momentum conservation equations(Appendix~\ref{app:quad:cons}), the D2Q9 model
fails to capture the Navier-Stokes viscous stress due to $O({\rm Ma}^3)$ deviations (Appendix~\ref{app:quad:CE}). 
Appendix~\ref{app:axi} presents analytical solutions of the Navier-Stokes equations for sound and shear
wave problems. Appendix~\ref{app:rot} describes the procedure for mapping scalar and vector quantities between
different spherical grids. Lastly, Appendix~\ref{app:Euler} discusses a solver for the inviscid Euler equations, used
as a reference for validating our method.

\section{The Navier-Stokes Equations on the sphere}\label{sec:hydro}

In this section, we provide a self-contained derivation of the Navier-Stokes equations governing the dynamics of a viscous, 
isothermal fluid on the surface of a sphere. In Subsec.~\ref{sec:hydro:vielb}, 
we introduce the notation and the vielbein formalism, in differential geometry language, specializing
it to the case of the sphere in Subsec.~\ref{sec:hydro:sph}. 
The reader familiar with this formalism may wish to jump directly to Subsec.~\ref{sec:hydro:eqs},
where we derive the equations of motion for a fluid on the surface of the sphere. 

We then further restrict the equations to axisymmetric configurations in Subsection~\ref{sec:hydro:axi}, 
enabling the derivation of new analytical solutions for the dynamics of sound and shear waves. 
These solutions, presented in Appendix~\ref{app:axi}, are a contribution of this work and serve as benchmarks 
for validating our numerical scheme, as discussed in Section~\ref{sec:numerics}.

\subsection{Preliminaries: The vielbein formalism}\label{sec:hydro:vielb}
Vectors and tensors in curved spaces are generally treated by employing curvilinear coordinates and differential geometry. 
These allow us to define fundamental notions such as distances, derivatives and integrals in a non-flat geometry, which are necessary to describe physical phenomena.
First, a curvilinear coordinate system that parametrizes the curved space is chosen. 
Vector fields, such as the velocity field $\vb{u{(x)}}$, can be expressed in 
a curvilinear coordinate system
as:
\begin{equation}\label{eq:vector_def_coord_space}
    \vb{u{(x)}} = u^a{(q^b)} \vb*{\partial}_a ,
\end{equation}
where $q^b$ represent the curvilinear coordinates chosen to parametrize the surface, 
$u^a{(q^b)}$ represent the velocity field components and $\vb*{\partial}_a$ represent the basis vectors.
Hereafter, we adopt Einstein's summation convention over repeated indexes.
The spherical surface can be seen as a two-dimensional manifold embedded in a three-dimensional space, so that $a, b = \{1,2\}$.

The squared norm of the velocity field $\vb{u}$ is given by:
\begin{equation}\label{eq:norm_def_coord_space}
    \vb{u}^2 = g_{ab} u^a u^b.
\end{equation}
The metric tensor $g_{ab}$ can be read off from the line element ${\dd s}^2 = dx^2 + dy^2 + dz^2$,
which is expressed with respect to the coordinates $\{q^a\}$ parametrizing the surface via
\begin{equation}\label{eq:line_element_metric_tensor_generic}
    {\dd s}^2 = g_{ab} \dd q^a \dd q^b , \qquad g_{ab} = \delta_{ij} \pdv{x^i}{q^a} \pdv{x^j}{q^b},
\end{equation}
with $i,j = \{1,2,3\}$ referring to the Cartesian coordinates and $a,b$ referring
to the curvilinear coordinates.

Next, we define a set of vector fields $\vb{e}_{\hat{a}} = e^{a}_{\hat{a}} \vb*{\partial}_a $ 
and use it as an orthonormal basis to rewrite vector quantities
so that the velocity field becomes:
\begin{equation}\label{eq:vector_def_vielbein}
    \vb{u} = u^{\hat{a}} \vb*{e}_{\hat{a}} .
\end{equation}
The vielbein vector fields (frame) and its dual vielbein one-form fields (co-frame), 
defined by $\vb*{\omega}^{\hat{a}}=\omega^{\hat{a}}_{a} \vb*{\dd q}^a$, 
satisfy the following relations~\cite{ambrus-pre-2019}:
\begin{equation}\label{eq:vielbein_orthonorm}
    g_{ab} e^{a}_{\hat{a}} e^{b}_{\hat{b}} = \delta_{\hat{a}\hat{b}} , \quad
    \omega^{\hat{a}}_{a} e^{a}_{\hat{b}} = \delta^{\hat{a}}_{\hat{b}}, \quad 
    \omega^{\hat{a}}_{a} e^{b}_{\hat{a}} = \delta^{b}_{a} ,
\end{equation}
where hatted indices refer to the vielbein frame components and 
non-hatted indices to coordinate space components. 
It follows that the components of the vector field $\vb{u}$ can be expressed as 
\begin{equation}\label{eq:vielbein_vec_components}
    u^{\hat{a}} = \omega^{\hat{a}}_{a} u^a, \quad 
    u^a         = e^{a}_{\hat{a}} u^{\hat{a}} ,
\end{equation}
with squared norm
\begin{equation}\label{eq:norm_def_vielbein}
    \vb{u}^2 = \delta_{\hat{a} \hat{b}} u^{\hat{a}} u^{\hat{b}} .
\end{equation}

We now introduce the Cartan coefficients and the connection coefficients.
The Cartan coefficients $c\indices{_{\hat{a}\hat{b}}^{\hat{c}}}$ are defined as the contraction
between the co-frame one-form $\vb*{\omega}^{\hat{c}}$ and the commutator of the frame vector field
\begin{equation}
    [\vb*{e}_{\hat{a}},\vb*{e}_{\hat{b}}] 
    := 
    (e^{a}_{\hat{a}}\partial_{a}e^{c}_{\hat{b}} - e^{b}_{\hat{b}}\partial_{b}e^{c}_{\hat{a}})\vb*{\partial}_{c} 
    \equiv 
    c_{\hat{a}\hat{b}}{}^{\hat{c}} e_{\hat{c}}^c \boldsymbol{\partial}_c. 
\end{equation}
Therefore, the Cartan coefficients can be expressed as:
\begin{equation}\label{eq:Cartan_coefs_def} 
    c\indices{_{\hat{a}\hat{b}}^{\hat{c}}}:= %
    \omega^{\hat{c}}_{c}(e^{a}_{\hat{a}}\partial_{a}e^{c}_{\hat{b}} - e^{b}_{\hat{b}}\partial_{b}e^{c}_{\hat{a}}) .
\end{equation}
The connection coefficients $\Gamma\indices{^{\hat{d}}_{\hat{b} \hat{c}}}$ directly relate to the space curvature and the choice of curvilinear coordinates.
They are linked to the Cartan coefficients via:
\begin{equation}\label{eq:connection_coefs_def}
    \Gamma\indices{^{\hat{d}}_{\hat{b}\hat{c}}} = \delta^{\hat{a}\hat{d}} \Gamma_{\hat{a}\hat{b}\hat{c}} , \quad
    \Gamma_{\hat{a}\hat{b}\hat{c}} = \frac{1}{2} (c_{\hat{a}\hat{b}\hat{c}} + c_{\hat{a}\hat{c}\hat{b}} - c_{\hat{b}\hat{c}\hat{a}}).
\end{equation}

As usual, it is possible to lower and raise indices in the vielbein frame via the Kronecker delta 
$\delta_{\hat{a}\hat{b}}$ and its inverse $\delta^{\hat{a}\hat{b}}$, 
e.g. 
\begin{equation}
    c_{\hat{b}\hat{c}\hat{d}} 
    = 
    \delta_{\hat{a}\hat{d}} c\indices{_{\hat{b}\hat{c}}^{\hat{a}}} 
    = 
    c\indices{_{\hat{b}\hat{c}}^{\hat{d}}}.
\end{equation}

From the definition above, one can define the 
gradient of a scalar function $F$ in the vielbein formalism as:
\begin{equation}\label{eq:gradient_vielbein}
    \nabla_{\hat{a}}{F} = e^{a}_{\hat{a}} \partial_a {F} .
\end{equation}
Moreover, the covariant derivative of a vector field $\vb{u}$ is:
\begin{equation}\label{eq:cov_der_vec}
    \nabla_{\hat{b}} u^{\hat{a}} 
    = 
    e^{b}_{\hat{b}} \partial_b u^{\hat{a}} + \Gamma\indices{^{\hat{a}}_{\hat{c} \hat{b}}} u^{\hat{c}}, \  
    \nabla_{\hat{b}} u_{\hat{a}} 
    = 
    e^{b}_{\hat{b}} \partial_b u_{\hat{a}} - \Gamma\indices{^{\hat{c}}_{\hat{a} \hat{b}}} u_{\hat{c}} ,
\end{equation}
whereas for a rank-two tensor $\tau^{\hat{a}\hat{b}}$ we have:
\begin{equation}\label{eq:cov_der_covariant_tensor}
    \nabla_{\hat{c}} \tau^{\hat{a}\hat{b}} = e^{c}_{\hat{c}} \partial_c \tau^{\hat{a}\hat{b}}
    + \Gamma\indices{^{\hat{a}}_{\hat{d} \hat{c}}} \tau^{\hat{d}\hat{b}} 
    + \Gamma\indices{^{\hat{b}}_{\hat{d} \hat{c}}} \tau^{\hat{a}\hat{d}} .
\end{equation}
Finally, the divergence of a vector $\vb{u}$ is:
\begin{equation}
    \nabla_{\hat{a}} u^{\hat{a}} = \dfrac{1}{\sqrt{g}} \partial_a( \sqrt{g} e^{a}_{\hat{a}} u^{\hat{a}} ), \label{eq:covariant_div} 
\end{equation}
with $\sqrt{g} = \sqrt{\det (g_{ab})}$ being the square root of the determinant of the metric tensor.

\subsection{Vielbein formalism for spherical coordinates}\label{sec:hydro:sph}
\begin{figure}[tb]
    \centering
    \includegraphics[width=0.99\columnwidth]{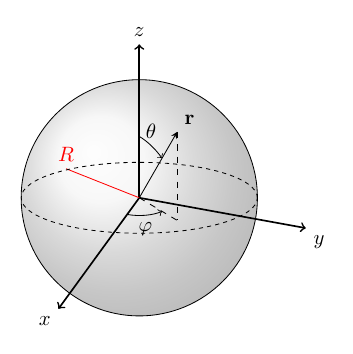}
    \caption{Spherical surface parametrized by the angles $\theta\in[0,\pi]$ and $\varphi\in[0,2\pi)$, with radius $R$. 
             The vector $\vb{r}$ marks the position vector for a generic point on the spherical surface.
            }\label{fig:sphere}
\end{figure}

We now specialize the formalism described above to the case of a spherical surface of radius $R$,
which we parametrize using the spherical polar coordinates $\theta\in [0,\pi]$ and $\varphi\in[0,2\pi)$.
The position vector $\mathbf{r} = x \mathbf{i} + y \mathbf{j} + z \mathbf{k}$ of a point on the 
spherical surface, shown in Fig.~\ref{fig:sphere}, 
is expressed in terms of the spherical coordinates via
\begin{equation}\label{eq:spherical_surface_parametrization}
    x = R\sin{\theta}\cos{\varphi}, \quad
    y = R\sin{\theta}\sin{\varphi}, \quad
    z = R\cos{\theta}.
\end{equation}
Substituting the above relations in Eq.~\eqref{eq:line_element_metric_tensor_generic} leads to
the following line element,
\begin{equation}\label{eq:line_element_spherical_surface}
    {\dd s}^2 = R^2{\dd \theta}^2 + R^2 \sin^2{\theta}{\dd \varphi}^2,
\end{equation}
which yields the metric tensor:
\begin{equation}\label{eq:metric_tensor_sphere}
	g_{ab} = 
	    \begin{pmatrix}
	        R^2 & 0 \\
	        0 & R^2 \sin^2{\theta}
	    \end{pmatrix} , \quad 
    \sqrt{g} = R^2 \sin{\theta}.
\end{equation}

The diagonal vielbein fields that satisfy the orthonormality relations in Eqs.~\eqref{eq:vielbein_orthonorm} are defined by
\begin{align}\label{eq:vielbein_frame_sphere}
    e^{\theta}_{\hat{\theta}} = \dfrac{1}{R},
    \quad
    e^{\varphi}_{\hat{\varphi}} = \dfrac{1}{R\sin{\theta}}, \quad 
    e^{\theta}_{\hat{\varphi}}=e^{\varphi}_{\hat{\theta}}=0.
\end{align}
The co-frame of vielbein 1-forms satisfying Eqs.~\eqref{eq:vielbein_orthonorm} 
have components
\begin{equation}\label{eq:vielbein_coframe_sphere}
    \omega^{\hat{\theta}}_{\theta} = R,
    \quad
    \omega^{\hat{\varphi}}_{\varphi} = R\sin{\theta}, \quad 
    \omega^{\hat{\theta}}_\varphi = \omega^{\hat{\varphi}}_\theta = 0.
\end{equation}

The commutator of the basis vectors evaluates to
\begin{align}\label{eq:vielbein_frame_commutator_sphere}
        [\vb*{e}_{\hat{\theta}},\vb*{e}_{\hat{\varphi}}] = - [\vb*{e}_{\hat{\varphi}},\vb*{e}_{\hat{\theta}}] = - \dfrac{\cos{\theta}}{R\sin{\theta}}\vb*{e}_{\hat{\varphi}}.
\end{align}
Using Eq.~\eqref{eq:Cartan_coefs_def}, the non-vanishing Cartan coefficients read
\begin{equation}\label{eq:Cartan_coefs_sphere}
    c\indices{_{\hat{\theta}\hat{\varphi}}^{\hat{\varphi}}} = - c\indices{_{\hat{\varphi}\hat{\theta}}^{\hat{\varphi}}} = - \dfrac{\cos{\theta}}{R\sin{\theta}} .
\end{equation}
Substituting the above into Eq.~\eqref{eq:connection_coefs_def}, we find the only non-zero connection coefficients to be:
\begin{align}\label{eq:connection_coefs_sphere}
    \Gamma_{\hat{\theta}\hat{\varphi}\hat{\varphi}} = - \Gamma_{\hat{\varphi}\hat{\theta}\hat{\varphi}} &=  \pqty{ -\dfrac{\cos{\theta}}{R\sin{\theta}} } .
\end{align}
The divergence of a four-vector $u^{\hat{a}}$ can be obtained using Eq.~\eqref{eq:covariant_div}:
\begin{equation}\label{eq:div_sph}
 \nabla_{\hat{a}} u^{\hat{a}} = \frac{\partial_\theta(\sin\theta u^{\hat{\theta}})}{R \sin\theta} + \frac{\partial_\varphi u^{\hat{\varphi}}}{R \sin\theta}.
\end{equation}

\subsection{Hydrodynamics in spherical coordinates}\label{sec:hydro:eqs}

In this section, we derive the hydrodynamic equations for a fluid on a spherical 
surface. The starting point is given by the continuity equation and \ac{nse} in covariant vielbein formalism:
\begin{subequations}
\begin{align}
    \partial_t{\rho} + \nabla_{\hat{a}}(\rho u^{\hat{a}}) &= 0, \label{eq:continuity_covariant} \\
    \rho \cdv{u^{\hat{a}}}{t} - \nabla_{\hat{b}}{\tau^{\hat{a}\hat{b}}} &= 0, \label{eq:NSE_covariant}
\end{align}
\end{subequations}
where $\mathrm{D}/\mathrm{D}{t}:= \partial_t + u^{\hat{b}} \nabla_{\hat{b}}$ is the convective derivative, $\rho$ is the fluid density, and $u^{\hat{a}}$ is the fluid velocity. The stress tensor $\tau^{\hat{a}\hat{b}}$ of a viscous ideal gas is
\begin{equation}\label{eq:stress-tensor}
    \tau^{\hat{a}\hat{b}} = -p \delta^{\hat{a}\hat{b}} + \sigma^{\hat{a}\hat{b}},    
\end{equation}
with pressure $p = n k_B T$, in which $n = \rho / m$ is the number density, $m$ is the particle mass, $T$ is the gas temperature and $k_B$ is the Boltzmann constant. The viscous stress tensor $\sigma^{\hat{a}\hat{b}}$ for a two-dimensional manifold reads~\cite{busuioc-jfm-2020}:
\begin{equation}\label{eq:viscous_stress}
    \sigma^{\hat{a}\hat{b}} 
    = 
    \eta \pqty{ \nabla^{\hat{a}}u^{\hat{b}} 
    + \nabla^{\hat{b}}u^{\hat{a}} 
    - \delta^{\hat{a}\hat{b}} \nabla_{\hat{c}}u^{\hat{c}} } 
    + \zeta \delta^{\hat{a}\hat{b}} \nabla_{\hat{c}}u^{\hat{c}} ,
\end{equation}
with $\eta$ and $\zeta$ being the dynamic and bulk viscosities, respectively. 
Using Eq.~\eqref{eq:div_sph} for the divergence of the velocity field,
the components of the stress tensor \eqref{eq:stress-tensor} can be obtained as follows:
\begin{align}
    \tau^{\hat{\theta}\hat{\theta}} &= -p + \frac{\eta + \zeta}{R} \partial_{\theta} u^{\hat{\theta}} - \frac{\eta - \zeta}{R \sin\theta} (\cos\theta u^{\hat{\theta}} + \partial_\varphi u^{\hat{\varphi}}), \nonumber\\
    \tau^{\hat{\varphi}\hat{\varphi}} &= -p - \frac{\eta - \zeta}{R} \partial_\theta u^{\hat{\theta}} + \frac{\eta + \zeta}{R \sin\theta} (\cos{\theta} u^{\hat{\theta}} + \partial_{\varphi} u^{\hat{\varphi}}), \\
    \tau^{\hat{\theta}\hat{\varphi}} &= \frac{\eta}{R}\sin\theta \frac{\partial}{\partial\theta}\left(\frac{u^{\hat{\varphi}}}{\sin\theta}\right) + \frac{\eta \partial_\varphi u^{\hat{\theta}}}{R \sin\theta}, \nonumber
    \label{eq:stress-tensor-sph}
\end{align}
while $\tau^{\hat{\varphi}\hat{\theta}} = \tau^{\hat{\theta}\hat{\varphi}}$. The above expressions for the stress tensor in spherical coordinates are consistent with those reported in Ref.~\cite{landau-lifshitz-1987}.

The divergence of the stress tensor $\tau^{\hat{a}\hat{b}}$, required on the left-hand side of Eq.~\eqref{eq:NSE_covariant}, can be computed using
\begin{equation}\label{eq:div_stress_tensor_cov}
    \nabla_{\hat{b}}{\tau^{\hat{a}\hat{b}}} = 
    \dfrac{1}{\sqrt{g}} \partial_b \pqty{ \sqrt{g} e^{b}_{\hat{b}} \tau^{\hat{a}\hat{b}} } + 
    \Gamma^{\hat{a}}{}_{\hat{d}\hat{b}} \tau^{\hat{d}\hat{b}}, 
\end{equation}
hence:
\begin{align}
    \nabla_{\hat{b}}{\tau^{\hat{\theta}\hat{b}}} &= \dfrac{\partial_{\theta} \pqty{\sin{\theta} \tau^{\hat{\theta}\hat{\theta}}}}{R \sin{\theta}}  
     + \dfrac{\partial_{\varphi} \tau^{\hat{\theta}\hat{\varphi}}}{R \sin{\theta}} -\frac{\tau^{\hat{\varphi}\hat{\varphi}}}{R \tan\theta}, \\
    \nabla_{\hat{b}}{\tau^{\hat{\varphi}\hat{b}}} &= \dfrac{\partial_{\theta} \pqty{\sin{\theta} \tau^{\hat{\varphi}\hat{\theta}}}}{R \sin{\theta}}  
     + \dfrac{\partial_{\varphi} \tau^{\hat{\varphi}\hat{\varphi}}}{R \sin{\theta}} + 
     \frac{\tau^{\hat{\theta}\hat{\varphi}}}{R \tan\theta}.
     \label{eq:sph_dtau}
\end{align}

The convective derivative appearing on the left-hand side of Eq.~\eqref{eq:NSE_covariant} evaluates to
\begin{subequations}\label{eq:sph_Dudt}
\begin{align}
    \cdv{u^{\hat{\theta}}}{t} &= \partial_t {u^{\hat{\theta}}} 
    + \dfrac{u^{\hat{\theta}}}{R} \partial_{\theta} u^{\hat{\theta}} 
    + \dfrac{u^{\hat{\varphi}} \partial_{\varphi} u^{\hat{\theta}}}{R \sin{\theta}} 
    - \dfrac{u^{\hat{\varphi}} u^{\hat{\varphi}}}{R\tan{\theta}}, \label{eq:sph_Duthdt}\\  
    \cdv{u^{\hat{\varphi}}}{t} &= \partial_t {u^{\hat{\varphi}}} 
    + \dfrac{u^{\hat{\theta}}}{R} \partial_{\theta} u^{\hat{\varphi}} 
    + \dfrac{u^{\hat{\varphi}} \partial_{\varphi} u^{\hat{\varphi}}}{R \sin{\theta}}
    + \dfrac{u^{\hat{\theta}} u^{\hat{\varphi}}}{R\tan{\theta}}. \label{eq:sph_Duphdt}
\end{align}
\end{subequations}

We are now ready to give the explicit form of the covariant hydrodynamic equations \eqref{eq:NSE_covariant}. 
The continuity equation \eqref{eq:continuity_covariant} can be obtained by replacing $u^{\hat{a}} \to \rho u^{\hat{a}}$ in Eq.~\eqref{eq:div_sph}:
\begin{subequations}\label{eq:NSE_sph}
\begin{equation}\label{eq:NSE_sph_continuity}
    \partial_t{\rho} + \dfrac{\partial_{\theta}(\rho \sin{\theta} u^{\hat{\theta}})}{R \sin{\theta}}  + \dfrac{\partial_{\varphi}(\rho u^{\hat{\varphi}})}{R \sin{\theta}} = 0.
\end{equation}
Substituting the convective derivative from Eqs.~\eqref{eq:sph_Dudt} and the divergence of the stress tensor from Eqs.~\eqref{eq:sph_dtau}, a tedious calculation leads to the \ac{nse} without external forces:
\begin{multline}\label{eq:NSE_sph_uth}
    \rho \pqty{ \partial_t u^{\hat{\theta}} 
    + \dfrac{u^{\hat{\theta}}}{R} \partial_{\theta} u^{\hat{\theta}} 
    + \dfrac{u^{\hat{\varphi}} \partial_{\varphi} u^{\hat{\theta}}}{R \sin{\theta}}  - \dfrac{u_{\hat{\varphi}}^2}{R\tan{\theta}}  }  = -\frac{\partial_\theta p}{R} \\
    + \frac{\eta + \zeta}{R^2} \frac{\partial}{\partial \theta} \left[\frac{\partial_\theta(\sin\theta u^{\hat{\theta}})}{\sin\theta}\right] + \frac{2\eta}{R^2} u^{\hat{\theta}} + \frac{\eta \partial^2_\varphi u^{\hat{\theta}}}{R^2 \sin^2\theta} \\
    + \frac{\zeta}{R^2} \frac{\partial}{\partial \theta} \left(\frac{\partial_\varphi u^{\hat{\varphi}}}{\sin\theta}\right) - \frac{2\eta}{R^2} \cos\theta \partial_\varphi u^{\hat{\varphi}},
\end{multline}
\begin{multline}\label{eq:NSE_sph_uph}
    \rho \pqty{ \partial_t u^{\hat{\varphi}} 
    + \dfrac{u^{\hat{\theta}}}{R} \partial_{\theta} u^{\hat{\varphi}} 
    + \dfrac{u^{\hat{\varphi}} \partial_{\varphi} u^{\hat{\varphi}}}{R \sin{\theta}} + \dfrac{u^{\hat{\theta}} u^{\hat{\varphi}}}{R\tan{\theta}} }  =
    - \dfrac{\partial_{\varphi}p}{R\sin{\theta}}  \\
    + \frac{(\eta + \zeta) \partial^2_\varphi u^{\hat{\varphi}}}{R^2 \sin^2\theta} 
    + \frac{\zeta \partial_\theta(\sin\theta \partial_\varphi u^{\hat{\theta}})}{R^2 \sin^2 \theta}
    + \frac{2\eta \partial_\varphi u^{\hat{\theta}}}{R^2 \sin\theta \tan\theta}  \\ 
    + \frac{\eta}{R^2 \sin^2\theta} \frac{\partial}{\partial\theta} \left[\sin^3\theta \frac{\partial}{\partial\theta} \left(\frac{u^{\hat{\varphi}}}{\sin\theta}\right)\right],
\end{multline}
where we assumed that the dynamical and bulk viscosities, $\eta$ and $\zeta$, are coordinate-independent.
\end{subequations}

\subsection{Axisymmetric flows} \label{sec:hydro:axi}

For benchmarking purposes, we will consider the case when the flow configuration is axisymmetric. In this case, the derivatives with respect to $\varphi$ vanish and the \ac{nse}~\eqref{eq:NSE_sph} reduce to:
\begin{subequations}\label{eq:NSE_axi}
\begin{equation}\label{eq:axi_continuity}
 \partial_t \rho + \frac{\partial_\theta(\rho \sin\theta u^{\hat{\theta}})}{R \sin\theta} = 0,
\end{equation}
\begin{multline}\label{eq:axi_NSE_uth}
 \rho \pqty{ \partial_t u^{\hat{\theta}} 
    + \dfrac{u^{\hat{\theta}}}{R} \partial_{\theta} u^{\hat{\theta}} 
    - \dfrac{u_{\hat{\varphi}}^2}{R\tan{\theta}}  }  = -\frac{\partial_\theta p}{R} \\
    + \frac{\eta + \zeta}{R^2} \frac{\partial}{\partial \theta} \left[\frac{\partial_\theta(\sin\theta u^{\hat{\theta}})}{\sin\theta}\right] + \frac{2\eta}{R^2} u^{\hat{\theta}},
\end{multline}
\begin{multline}\label{eq:axi_NSE_uph}
    e \rho \pqty{ \partial_t u^{\hat{\varphi}} 
    + \dfrac{u^{\hat{\theta}}}{R} \partial_{\theta} u^{\hat{\varphi}} 
    + \dfrac{u^{\hat{\theta}} u^{\hat{\varphi}}}{R\tan{\theta}} }  =  \\
    \frac{\eta}{R^2 \sin^2\theta} \frac{\partial}{\partial\theta} \left[\sin^3\theta \frac{\partial}{\partial\theta} \left(\frac{u^{\hat{\varphi}}}{\sin\theta}\right)\right].
\end{multline}
\end{subequations}

\section{Numerical Method: vielbein lattice Boltzmann on the sphere}\label{sec:method}

In this section, we describe the steps required to define the discrete Boltzmann equation
in the vielbein formalism on the sphere.
The reader not concerned with the details can skip directly to Sec.~\ref{sec:method:summary}, where we give a summary of the method.
The starting point is given by the covariant Boltzmann equation~\cite{ambrus-pre-2019,busuioc-pre-2019}, which we consider in the absence of external forces:
\begin{equation}
    \pdv{f}{t} + \dfrac{1}{\sqrt{g}} \pdv{q^{b}} \pqty{ v^{\hat{a}} e^{b}_{\hat{a}}f \sqrt{g} } 
    - \Gamma^{\hat{a}}{}_{\hat{b}\hat{c}}
    \frac{\partial(v^{\hat{b}} v^{\hat{c}} f)}{\partial v^{\hat{a}}} = C[f], 
    \label{eq:Boltzmann_eq_covariant}
\end{equation}
where $f \equiv f(q^a, v^{\hat{b}}, t)$ is the single-particle distribution function, $v^{\hat{a}}$ are the vielbein components of the particle velocity and $C[f]$ is the collision term.
The third term on the left-hand side of Eq.~\eqref{eq:Boltzmann_eq_covariant} shows that the connection coefficients give rise to forcing terms, whose role is to enforce the propagation of fluid particles along the curved surface. They are thus related to the intrinsic curvature of the chosen geometry or the employment of curvilinear coordinates.

In the case of the spherical geometry described in Sec.~\ref{sec:hydro:sph}, Eq.~\eqref{eq:Boltzmann_eq_covariant} becomes:
\begin{multline}
        \pdv{f}{t} + \dfrac{v^{\hat{\theta}}}{R\sin{\theta}} \pdv{(f\sin{\theta})}{\theta} + \dfrac{v^{\hat{\varphi}}}{R\sin{\theta}} \pdv{f}{\varphi} \\
        + \dfrac{\cos{\theta}}{R\sin{\theta}}\bqty{v^{\hat{\varphi}}\pdv{(f v^{\hat{\varphi}})}{v^{\hat{\theta}}} - v^{\hat{\theta}}\pdv{(f v^{\hat{\varphi}})}{v^{\hat{\varphi}}}} = C[f].
        \label{eq:Boltzmann_eq_sphere}
\end{multline}
This is the equation that we need to discretize in order to derive the \ac{lbe} and construct our \ac{vlbm} for the simulation
of fluid flows on the surface of a sphere.

In the remainder of this section, we present all the steps required for the construction of the lattice Boltzmann algorithm for the solution of Eq.~\eqref{eq:Boltzmann_eq_sphere}. In Subsec.~\ref{sec:method:vel}, we discuss the discretization of the velocity space using the Gauss-Hermite quadrature. 
Subsec.~\ref{sec:method:BGK} presents the construction of the collision term, for which we employ the BGK model~\cite{bhatnagar-pr-1954}. 
Subsec.~\ref{sec:method:force} describes the computation of the derivatives with respect to the velocity, which are required to implement the curvature terms.
Subsections~\ref{sec:method:space} and \ref{sec:method:advection} present the space discretization and the flux-based, finite-difference advection scheme, while boundary conditions are discussed in Subsec.~\ref{sec:method:bcs}. 
Finally, Subsec.~\ref{sec:method:time} presents the time-stepping algorithm, for which we employ the Runge-Kutta explicit time stepping.
\subsection{Velocity discretization}\label{sec:method:vel}
One key aspect in the derivation of a \ac{lbm} is the discretization of the velocity space. 
To achieve this, we introduce the tensor \ac{hp}s, 
denoted as $\mathcal{H}^{(\ell)}_{\alpha_1,\dots,\alpha_{\ell}}(\vb{v})$ and defined as \cite{shan-jofm-2006,sofonea-pre-2018}:
\begin{equation}\label{eq:hermite_tensor_poly}
    \mathcal{H}^{(\ell)}_{\alpha_1,\dots,\alpha_{\ell}}(\vb{v}) = 
    \dfrac{(-1)^\ell }{\omega(\vb{v})} 
    \frac{\partial^\ell \omega(\vb{v})}{\partial v_{\alpha_1} \cdots \partial v_{\alpha_\ell}}, \quad 
    \omega(\vb{v}) = \dfrac{e^{-\mathbf{v}^2/2}}{(2\pi)^{D/2}}.
\end{equation}
For a $D$-dimensional space, the indices satisfy $1 \le \alpha_{1},\dots,\alpha_{\ell} \le D$, 
and an expansion up to order $Q$ includes polynomials of orders $0 \le \ell < Q$. 
The first four \ac{hp}s are:
\begin{gather}\label{eq:hermite_polynomials}
    \mathcal{H}^{(0)}(\vb{v}) = 1, \quad 
    \mathcal{H}^{(1)}_{\alpha}(\vb{v}) = v_{\alpha}, \quad 
    \mathcal{H}^{(2)}_{\alpha \beta}(\vb{v}) = v_{\alpha} v_{\beta}-\delta_{\alpha \beta},\nonumber\\
    \mathcal{H}^{(3)}_{\alpha \beta \gamma}(\vb{v}) = v_{\alpha} v_{\beta} v_{\gamma}-\pqty{\delta_{\alpha \beta} v_{\gamma}+\delta_{\beta \gamma} v_{\alpha}+\delta_{\gamma \alpha} v_{\beta}} .
\end{gather}
Following a Gauss-Hermite quadrature rule, the roots of the one-dimensional \ac{hp} $H_Q(v_{k_\alpha})$ ($1 \le k_\alpha \le Q$) 
define the Cartesian components $v_{k_\alpha}$ of the discrete velocity set along axis $\alpha$. 
The vielbein formalism ensures that these components remain valid even in non-Cartesian geometries.

The quadrature weights for order $Q$ along axis $\alpha$ are given by:
\begin{equation}\label{eq:GH_weights}
    w_{k_\alpha} = \dfrac{Q!}{H^2_{Q+1}(v_{k_\alpha})} .
\end{equation}
The overall quadrature weights are $w_{\vb{k}}=w_{k_{\alpha_1}}\dots w_{k_{\alpha_\ell}}$.

A Gauss-Hermite quadrature of order $Q$ \cite{kruger-book-2017,sofonea-pre-2018} exactly recovers all moments up to order 
$Q - 1$ of the equilibrium distribution function $f^\mathrm{eq}$, by projecting $f^{\mathrm{eq}}$ 
onto the space of Hermite polynomials (see Subsec.~\ref{sec:method:BGK}). 
Therefore, conserved moments can be computed exactly via discrete summation over $f_{\vb{k}}$:
\begin{equation}\label{eq:f_moments_sum}
    n        = \sum_{\vb{k}} f_{\vb{k}}, \quad
    n \vb{u} = \sum_{\vb{k}} f_{\vb{k}} \vb{v_k} ,
\end{equation}
with the discrete distribution function $f_{\vb{k}}$ related to the Boltzmann distribution $f$ via:
\begin{equation}\label{eq:f_disc}
    f_{\vb{k}} = \dfrac{w_{\vb{k}}}{\omega(\vb{v_k})} f(\vb{v_k}) .
\end{equation}
The exact recovery of the first three moments of the distribution function is essential for any kinetic model that aims to capture the correct hydrodynamic equations. The transition from the mesoscopic kinetic description to the macroscopic \ac{nse} framework is achieved via a Chapman-Enskog expansion in the Knudsen number \cite{kruger-book-2017} (see Appendix~\ref{app:quad:CE}).

\subsection{Collision term discretization}\label{sec:method:BGK}
In this paper, we consider the  Bhatnagar-Gross-Krook (BGK) model for the collision operator,
\begin{equation}
    C[f] = -\frac{1}{\tau}(f - f^{\mathrm{eq}}),
\end{equation}
where $\tau$ is the relaxation time, and $f^{\mathrm{eq}}$ is the local equilibrium distribution, 
modeled using the Maxwell-Boltzmann distribution,
\begin{equation}
    f^{\mathrm{eq}} = n \left(\frac{m}{2\pi k_B T}\right)^{D/2} \exp\left[-\frac{m(\mathbf{v} - \mathbf{u})^2}{2 k_B T}\right].
\end{equation}
In what follows, we consider isothermal flows at $T = T_0$, 
and adopt a non-dimensionalization of velocities based on the thermal reference speed
$c_{\rm ref} = \sqrt{k_B T / m}$, such that the expression for the local equilibrium simplifies down to 
\begin{equation}
    f^{\rm eq} = \frac{n}{(2\pi)^{D/2}} e^{-(\mathbf{v} - \mathbf{u})^2 / 2}.
\end{equation}

In order to discretize the collision operator, we consider an expansion of the
equilibrium distribution in terms of \ac{hp} \cite{shan-jofm-2006,sofonea-pre-2018}:
\begin{equation}\label{eq:feq_series}
	f^{\mathrm{eq}}(\vb{x},\vb{v},t)= \omega(\vb{v}) \sum_{\ell=0}^{\infty} \frac{1}{\ell!} 
    a^{\mathrm{eq},(\ell)}_{\alpha_{1},\dots,\alpha_{\ell}}(\vb{x}, t) 
      \mathcal{H}^{(\ell)}_{\alpha_{1},\dots,\alpha_{\ell}}(\vb{v}),
\end{equation}
where a summation over the repeated Greek indices is implied,
and the expansion coefficients are defined as:
\begin{equation}
    a^{\mathrm{eq},(\ell)}_{\alpha_{1}, \ldots, \alpha_{n}}(\vb{x}, t)
    =
    \int \dd{\vb{v}} f^{\mathrm{eq}}(\vb{x}, \vb{v}, t) \mathcal{H}^{(\ell)}_{\alpha_{1},\dots,\alpha_{\ell}}(\vb{v}) .
\end{equation}
It follows that the first four coefficients are given by:
\begin{gather}\label{eq:feq_coefs}
    a^{\mathrm{eq},(0)} = n, \quad
    a^{\mathrm{eq},(1)}_{\alpha} = n u_{\alpha}, \nonumber
    \\
    a^{\mathrm{eq},(2)}_{\alpha \beta } = n u_{\alpha} u_{\beta}, \quad
    a^{\mathrm{eq},(3)}_{\alpha \beta \gamma} = n u_{\alpha} u_{\beta} u_{\gamma} .
\end{gather}

Since the expansion coefficients coincide with the hydrodynamic quantities of interest,
it directly follows that considering an expansion of $f^{\mathrm{eq}}$ truncated at the order $N=Q-1$
\begin{equation}\label{eq:feq_trunc}
    f^{Q,\mathrm{eq}}(\vb{x}, \vb{v}, t) = \omega(\vb{v}) \sum_{\ell=0}^{Q-1} \frac{1}{\ell!} 
    a^{\mathrm{eq},(\ell)}_{\alpha_{1},\dots,\alpha_{\ell}}
      \mathcal{H}^{(\ell)}_{\alpha_{1},\dots,\alpha_{\ell}}(\vb{v}) ,
\end{equation}
all the moments of $f^{\rm eq}$ up to order $N = Q - 1$ are preserved exactly.

Next, following the discretization of the velocity space, the continuous distribution $f^{\rm eq}(\vb{x}, \vb{v}, t)$ 
is replaced by a discrete set of distributions $f^{\mathrm{eq}}_{\vb{k}}(\vb{x}, t)$,
which are related to the original continuous distribution as shown in Eq.~\eqref{eq:f_disc}, that is,
\begin{equation}\label{eq:feq_disc}
 f^{\mathrm{eq}}_{\vb{k}}(\vb{x}, t) = \frac{w_{\vb{k}}}{\omega(\vb{v}_{\vb{k})}} f^{\rm eq}(\vb{x}, \vb{v}_{\vb{k}}, t).
\end{equation}
In conclusion, the truncated discrete equilibrium distribution can be expressed as:
\begin{equation}\label{eq:feq_discrete}
    f^{Q,\mathrm{eq}}_{\vb{k}}(\vb{x}, t) = \omega(\vb{v}) \sum_{\ell=0}^{Q-1} \frac{1}{\ell!} 
    a^{\mathrm{eq},(\ell)}_{\alpha_{1},\dots,\alpha_{\ell}}
      \mathcal{H}^{(\ell)}_{\alpha_{1},\dots,\alpha_{\ell}}(\vb{v_k}) .
\end{equation}
Full expressions for the cases $Q=3$ and $Q=4$ can be found in Appendix~\ref{app:quad}.

\subsection{Velocity gradients discretization}\label{sec:method:force}
We now discuss the strategy for computing the velocity gradients $\pdv*{f}{v^{\hat{\theta}}}$ and $\pdv*{(fv^{\hat{\varphi}})}{v^{\hat{\varphi}}}$ appearing in Eq.~\eqref{eq:Boltzmann_eq_sphere},
by projecting these derivatives onto the space of the corresponding one-dimensional Hermite polynomials.
First, we consider the Hermite expansion of $f(v^{\hat{\theta}}, v^{\hat{\varphi}})$ with respect to either $v^{\hat{\theta}}$ or $v^{\hat{\varphi}}$:
\begin{align}\label{eq:f_series}
	f &= \omega(v^{\hat{\theta}}) \sum_{\ell=0}^{\infty} \frac{1}{\ell!} 
    \mathcal{F}^\theta_{\ell}(v^{\hat{\varphi}}) 
    H_\ell(v^{\hat{\theta}}) \notag \\ 
    &= \omega(v^{\hat{\varphi}}) \sum_{\ell=0}^{\infty} \frac{1}{\ell!} 
    \mathcal{F}^\varphi_{\ell}(v^{\hat{\theta}}) 
    H_\ell(v^{\hat{\varphi}}).
\end{align}
The velocity derivatives of $f$ can be taken using the following relations:
\begin{subequations}
\begin{align}
 \frac{\partial[\omega(v^{\hat{1}}) H_\ell(v^{\hat{1}})]}{\partial v^{\hat{1}}} &= - \omega(v^{\hat{1}}) H_{\ell + 1}(v^{\hat{1}}), \\
 \frac{\partial[\omega(v^{\hat{1}}) v^{\hat{1}} H_\ell(v^{\hat{1}})]}{\partial v^{\hat{1}}} &= -\omega(v^{\hat{1}}) [\ell H_\ell(v^{\hat{1}}) \nonumber\\
 &+ (\ell + 1)(\ell + 2) H_{\ell + 2}(v^{\hat{1}})].
\end{align}
\end{subequations}
This leads to the following results:
\begin{subequations}
\begin{align}
 \frac{\partial f}{\partial v^{\hat{1}}} &= -\omega(v^{\hat{1}}) \sum_{\ell=0}^{\infty} \frac{\ell}{\ell!} 
    \mathcal{F}_{\ell - 1}(v^{\hat{2}}) 
    H_{\ell}(v^{\hat{1}}), \\
 \frac{\partial (f v^{\hat{1}})}{\partial v^{\hat{1}}} &= -\omega(v^{\hat{1}}) \sum_{\ell=0}^{\infty} \frac{1}{\ell!} 
    [\ell \mathcal{F}_{\ell} + \ell(\ell - 1) \mathcal{F}_{\ell - 2}] H_{\ell}(v^{\hat{1}}).
\end{align}
\end{subequations}

The expansion coefficients $\mathcal{F}^\theta_{\ell}(v^{\hat{\varphi}})$ and $\mathcal{F}^{\varphi}_{\ell}(v^{\hat{\theta}})$ can be evaluated by exploiting the orthogonality of the Hermite polynomials,
\begin{equation}\label{eq:hermite_orthogonality}
    \int^{\infty}_{-\infty} \dd{x} \omega(x) H_\ell(x) H_{\ell'}(x) = \ell! \delta_{\ell\ell'},
    \ \omega(x)=\dfrac{ e^{-x^2/2} }{ \sqrt{2\pi} },
\end{equation}
giving 
\begin{align}
    \mathcal{F}^\theta_{\ell}(v^{\hat{\varphi}}) &= 
    \int_{-\infty}^\infty dv^{\hat{\theta}} \, f \, H_\ell(v^{\hat{\theta}}), \notag \\
    \mathcal{F}^\varphi_{\ell}(v^{\hat{\theta}}) &= 
    \int_{-\infty}^\infty dv^{\hat{\varphi}} \, f \, H_\ell(v^{\hat{\varphi}}).
    \label{eq:F_l}
\end{align}
After the velocity space discretization, $\mathcal{F}^\theta_{\ell}(v^{\hat{\varphi}})$ and $\mathcal{F}^\varphi_{\ell}(v^{\hat{\theta}})$ are replaced by
\begin{align}
    \mathcal{F}^\theta_{\ell,k_{\varphi}} &= \sum_{k_\theta=1}^{Q} H_\ell(v_{k_\theta}) f_{k_\theta,k_\varphi}, &
    \mathcal{F}^\varphi_{\ell,k_{\theta}} &= \sum_{k_\varphi=1}^{Q} H_\ell(v_{k_\varphi}) f_{k_\theta,k_\varphi}.
\end{align}
In an analogue way, we project and discretize the velocity gradients of $f$:
\begin{subequations}\label{eq:df}
\begin{align}
    \pqty{ \pdv{f}{v^{\hat{1}}}}_{\vb{k}}   &= \sum_{k'_1=1}^{Q} \mathcal{K}^H_{k_1,k'_1} f_{k'_1,k_2} \label{eq:dfdv}\\
    \pqty{ \pdv{(f v^{\hat{1}})}{v^{\hat{1}}} }_{\vb{k}} &= \sum_{k'_1=1}^{Q} \widetilde{\mathcal{K}}^H_{k_1,k'_1} f_{k'_1,k_2} ,
    \label{eq:dvfdv}
\end{align}
\end{subequations}
where the kernels $\mathcal{K}^H_{k_\theta,k'_\theta}$, $\widetilde{\mathcal{K}}^H_{k_\varphi,k'_\varphi}$ have the following expressions \cite{toschi-springer-2019, ambrus-pre-2019}:
\begin{subequations}\label{eq:kernels}
\begin{align}
    \mathcal{K}^H_{k,k'} &= - w_k \sum_{\ell=0}^{Q-1} \dfrac{1}{\ell!} H_{\ell+1}(v_k) H_{\ell}(v_{k'}), \label{eq:KH} \\ 
    \widetilde{\mathcal{K}}^H_{k,k'} &= - w_k \sum_{\ell=0}^{Q-1} \dfrac{1}{\ell!} H_{\ell+1}(v_k) \bqty{ H_{\ell+1}(v_{k'}) + \ell H_{\ell-1}(v_{k'}) }. \label{eq:KtildeH}
\end{align}
\end{subequations}

The explicit expressions of the objects introduced in this section for the quadrature orders $Q=3$ and $Q=4$ are found in Appendix~\ref{app:quad}.

\subsection{Space discretization}\label{sec:method:space}
The discretization of the velocity space via the Gauss-Hermite quadrature procedure seen above leads to a discrete version of the Boltzmann equation which, for the sphere, corresponds to:
\begin{multline}\label{eq:Boltzmann_eq_sphere_discrete}
        \pdv{f_{\vb{k}}}{t} + \dfrac{1}{R\sin{\theta}} \bqty{ \pdv{ (f_{\vb{k}} v_{k_\theta} \sin{\theta}) }{\theta} + \pdv{ (f_{\vb{k}} v_{k_\varphi}) }{\varphi} } \\
        + \dfrac{\cos{\theta}}{R\sin{\theta}} \bqty{ v^2_{k_\varphi} \pqty{ \pdv{f}{v^{\hat{\theta}}} }_{\vb{k}} - v_{k_\theta} \pqty{ \pdv{ (f v^{ \hat{\varphi}}) }{ v^{\hat{\varphi}}} }_{\vb{k}} } \\
        = -\dfrac{1}{\tau} \bqty{ f_{\vb{k}} - f^{\mathrm{eq}}_{\vb{k}} }.
\end{multline}
We now discretize the space via an equidistant grid along the $\theta$ and $\varphi$ coordinates. Since $\theta \in [0,\pi]$ and $\varphi \in [0,2\pi)$,
we take $N_\theta$ points along $\theta$ and $N_\varphi$ points along $\varphi$ such that:
\begin{subequations}\label{eq:discrete_coord_sphere}
\begin{align}
    \theta_s &= \frac{\pi}{N_\theta}(s-1/2) \qquad (1 \le s \le N_\theta), \\
    \varphi_q &= \frac{2\pi}{N_\varphi}(q-1/2) \qquad (1 \le q \le N_\varphi) .    
\end{align}
\end{subequations}
In this two-dimensional coordinate space, each cell $(s,q)$ has four interfaces which we write as $(s+1/2,q), (s-1/2,q), (s,q+1/2)$ and $(s,q-1/2)$. For simplicity, we will assume here equidistant cells along both $\theta$ and $\varphi$, thus
 $\var{\theta}=\pi/N_\theta$ and $\var{\varphi}=2\pi/N_\varphi$, respectively. Note, however, that in general it is desirable to consider the more general case where a non-uniform discretization is employed along $\theta$ to avoid accumulation of grid points in the proximity of the poles.
\subsection{Advection}\label{sec:method:advection}
To complete the spatial discretization of the method, we write the advection term appearing in the parentheses on the first line of \eqref{eq:Boltzmann_eq_sphere_discrete} as:
\begin{multline}\label{eq:discrete_advection_sphere}
    \Bqty{ \dfrac{1}{R \sin{\theta}} \bqty{ \pdv{\theta}( f_{\vb{k}} v_{k_\theta} \sin{\theta} ) + \pdv{\varphi} ( f_{\vb{k}} v_{k_\varphi} ) } }_{s,q} = \\
    \dfrac{1}{R \sin{\theta_s}} \bqty{ 
    \dfrac{\mathcal{F}^{\theta}_{s + \frac{1}{2},q}  - \mathcal{F}^{\theta}_{s - \frac{1}{2}, q }}{\delta \theta} + 
    \dfrac{\mathcal{F}^{\varphi}_{s, q + \frac{1}{2}} - \mathcal{F}^{\varphi}_{s, q - \frac{1}{2}}}{\delta \varphi}
    }.
\end{multline}
Here, $\mathcal{F}$ is used to represent the fluxes evaluated at the interfaces between adjacent cells (hence the $s\pm1/2$ and $q\pm1/2$ at the subscript) so that, for instance, $\mathcal{F}^{\theta}_{s + 1/2, q}$ is the flux of $f$ advected with velocity $v_{k_\theta}$ through the interface between the cell centered in $(s,q)$ and that centered in $(s+1,q)$ as in Ref.~\cite{busuioc-cm-2020}.

The choice of a \ac{weno5} scheme for the advection part of the Boltzmann equation was motivated by its high precision \cite{hejranfar-pre-2017}, as well as its shock-capturing properties \cite{jiang-jcp-1996,gan-pre-2011}. However, lower-order schemes may also be used, giving some advantages in terms of computational cost. To improve the accuracy near the poles, we support these lower-order models with a different definition of the fluxes in these areas corresponding to \cite{falle-mnras-1996}:
\begin{multline}\label{eq:Komissarov_scheme}
\hspace{-12pt}    \bqty{\dfrac{1}{R \sin{\theta}} \pdv{\theta}( f_{\vb{k}} v_{k_\theta} \sin{\theta} ) }_{s,q}  = 
  - \bqty{\dfrac{1}{R} \pdv{\cos{\theta}} ( f_{\vb{k}} v_{k_\theta} \sin{\theta} ) }_{s,q} \\ 
    \simeq 
    \dfrac{1}{R} \bqty{ 
    \dfrac{\mathcal{F}^{\theta}_{s + \frac{1}{2}, q} \sin{\theta_{s + \frac{1}{2}}} - \mathcal{F}^{\theta}_{s - \frac{1}{2}, q} \sin{\theta_{s - \frac{1}{2}}}}{\cos{{\theta_{s - \frac{1}{2}}}} - \cos{\theta_{s + \frac{1}{2}}}} }.
\end{multline}
Here, the $\sin{\theta}$ terms on the right-hand-side represent:
\begin{subequations}\label{eq:Komissarov_sin_cos}
\begin{align}
    \sin{\theta_{s + \frac{1}{2}}} &= \sin\pqty{ (s+1) \var{\theta}}, & 
    \sin{\theta_{s - \frac{1}{2}}} &= \sin\pqty{  s    \var{\theta}},  \\
    \cos{\theta_{s + \frac{1}{2}}} &= \cos\pqty{ (s+1) \var{\theta}}, & 
     \cos{\theta_{s - \frac{1}{2}}} &= \cos\pqty{  s    \var{\theta}}.
\end{align}
\end{subequations}
Here and in what follows, we refer to the approach in Eqs.~\eqref{eq:Komissarov_scheme}--\eqref{eq:Komissarov_sin_cos} as the ``Komissarov'' scheme. A known drawback of the Komissarov scheme is that, due to the addition of the geometrical factors of $\sin\theta_{s \pm \frac{1}{2}}$ next to the actual flux functions, the method cannot exceed second-order accuracy \cite{falle-mnras-1996,busuioc-pre-2019}.

We implement the fluxes in an upwind-biased manner, using a selection of schemes: the first-order (U1), second-order (U2) and third-order (U3) upwind schemes, as well as the \ac{weno5} scheme. Taking for definiteness the flux $\mathcal{F}^{\theta}_{s+\frac{1}{2}, q}$ for the case when the advection velocity $V$ is positive, with $V = v_{k_\theta} \sin\theta$ for Eq.~\eqref{eq:discrete_advection_sphere} and $V = v_{k_\theta}$ for the Komissarov scheme in Eq.~\eqref{eq:Komissarov_scheme}, the fluxes for the upwind schemes are:
\begin{subequations}\label{eq:fluxes_U1-3}
\begin{align}
 \text{U1}: 
 \mathcal{F}^{\theta}_{s+\frac{1}{2}, q} &= V f_{s,q}, \\
\text{U2}: 
 \mathcal{F}^{\theta}_{s+\frac{1}{2}, q} &= V\left(\frac{3}{2} f_{s,q} - \frac{1}{2} f_{s-1,q}\right), \\
 \text{U3}: 
 \mathcal{F}^{\theta}_{s+\frac{1}{2}, q} &= V \left(\frac{1}{3} f_{s+1,q} + \frac{5}{6} f_{s,q} - \frac{1}{6} f_{s-1,q}\right).
\end{align}
\end{subequations}
In the case of the WENO-5 scheme, we write $\mathcal{F}_{s+1/2} \equiv \mathcal{F}^{\theta}_{s+\frac{1}{2}, q}$ and henceforth drop the $\theta$ superscript, as well as the $q$ index. The flux is computed as an average over three stencils, $\mathcal{F}_{s+1/2} = \sum_{i = 1}^3 \overline{\omega}_i \mathcal{F}^i_{s+\frac{1}{2}}$, with interpolating functions
\begin{subequations}\label{eq:fluxes_W5}
\begin{align}
 \mathcal{F}^1_{s+\frac{1}{2}} &= V\left(\frac{1}{3} f_{s-2} - \frac{7}{6} f_{s-1} + \frac{11}{6} f_s\right), \\
 \mathcal{F}^2_{s+\frac{1}{2}} &= V\left(-\frac{1}{6} f_{s-1} + \frac{5}{6} f_s + \frac{1}{3} f_{s+1}\right), \\
 \mathcal{F}^3_{s+\frac{1}{2}} &= V\left(\frac{1}{3} f_s + \frac{5}{6} f_{s+1} - \frac{1}{6} f_{s+2}\right).
\end{align}
\end{subequations}
The weighing factors $\overline{\omega}_i = \tilde{\omega}_i / (\sum_{j = 1}^3 \tilde{\omega}_j)$, with $\tilde{\omega}_j = \delta_j / \sigma_j^2$, are given in terms of the ideal weights,
\begin{equation}\label{eq:WENO5_delta}
 \delta_1 = \frac{1}{10}, \quad \delta_2 = \frac{6}{10}, \quad \delta_3 = \frac{3}{10},
\end{equation}
as well as the smoothness functions
\begin{subequations}\label{eq:WENO5_sigma}
\begin{align}
 \sigma_1 &= \frac{13}{12}(f_{s-2} - 2f_{s-1} + f_s)^2 + \frac{1}{4}(f_{s-2} - 4f_{s-1} + 3f_s)^2, \\
 \sigma_2 &= \frac{13}{12}(f_{s-1} - 2f_s + f_{s+1})^2 + \frac{1}{4}(f_{s-1} - f_{s+1})^2, \\
 \sigma_3 &= \frac{13}{12}(f_s - 2f_{s+1} + f_{s+2})^2 + \frac{1}{4}(3f_s - 4f_{s+1} + f_{s+2})^2.
\end{align}
\end{subequations}

\subsection{Boundary Conditions}\label{sec:method:bcs}

\begin{figure}[t]
    \centering
    \includegraphics[width=0.99\columnwidth]{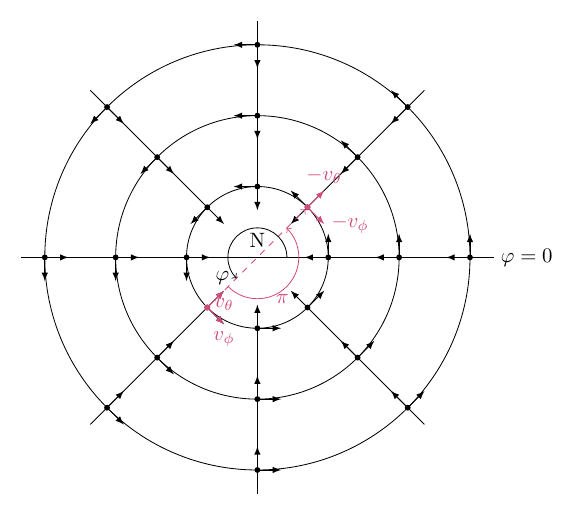}
    \caption{North pole view (with the letter N indicating the pole) of the boundaries defined by Eq.~\eqref{eq:bcs_theta} for 
             $\theta=0$ in the case of positive velocity. The black arrows represent the advection velocities at different grid points. 
             In red we highlight the components of a velocity vector traveling towards the north pole.
             Notice that the ``translated'' advection velocities have the opposite sign compared to 
             those of the population residing at the same node (black arrows).
            }\label{fig:boundaries_theta=0}
\end{figure}

For the implementation of the fluxes discussed in the previous subsection, the distribution function 
values to the left and/or right of the computation point must be utilized. 
Consequently, it is necessary to define the distribution function beyond the limits of the grid
$1 \le s \le N_\theta$, $1 \le q \le N_\varphi$. Since the sphere is a periodic manifold, 
the distribution function outside this grid can be mapped within it. 

This is straightforward for boundary conditions in the azimuthal ($\varphi$) coordinate:
\begin{subequations}\label{eq:bcs_phi}
\begin{align}
 f(\theta_s, \varphi_{N_\varphi + q}; \mathbf{v}) &= f(\theta_s, \varphi_{q}; \mathbf{v}), \\ 
 f(\theta_s, \varphi_{-q}; \mathbf{v}) &= f(\theta_s, \varphi_{N_\varphi - q}; \mathbf{v}).
\end{align}
\end{subequations}

For the polar angle ($\theta$), the boundary conditions require a more careful treatment. 
To illustrate this, let us consider a trajectory reaching the north pole at a fixed
azimuthal angle $\varphi$ (cf.~Fig.~\ref{fig:boundaries_theta=0}). 
As it crosses to the opposite side, the point with coordinates $(-\theta, \varphi)$ is mapped to $(\theta, \varphi \pm \pi)$, ensuring that $\varphi$ remains within the domain $[0,2\pi)$. 
Additionally, a velocity vector with components $(v^{\hat{\theta}}, v^{\hat{\varphi}})$ 
undergoes inversion upon crossing the pole.
Therefore, we formulate the polar boundary conditions as
\begin{subequations}\label{eq:bcs_theta}
    \begin{align}
     f(-\delta\theta,\varphi; \mathbf{v}) &= f(\delta\theta,\varphi \pm \pi; -\mathbf{v}), \\
     f(\pi + \delta\theta,\varphi; \mathbf{v}) &= f(\pi - \delta\theta,\varphi \pm \pi; -\mathbf{v}),
    \end{align}
\end{subequations}
where we choose $\delta \theta > 0$ for definiteness.

\subsection{Time stepping}\label{sec:method:time}

The introduction of finite difference schemes for the advection makes it necessary to employ finite difference schemes also for the time-marching part of the algorithm. In this work, a total variation diminishing third-order Runge-Kutta algorithm was used to approximate $\partial_t{f}$ at step $l$ \cite{toschi-springer-2019,busuioc-pre-2019}:
\begin{subequations}
\begin{align}
    f_{l}^{(1)} &= f_l + \var{t} \mathcal{L}{[f_l]} \\ 
    f_{l}^{(2)} &= \dfrac{3}{4} f_{l} + \dfrac{1}{4} f_{l}^{(1)} + \dfrac{1}{4} \var{t} \mathcal{L}{[f_{l}^{(1)}]} \\
    f_{l+1}     &= \dfrac{1}{3} f_{l}       + \dfrac{2}{3} f_{l}^{(2)} + \dfrac{2}{3} \var{t} \mathcal{L}{[f_{l}^{(2)}]},
\end{align}
\end{subequations}
assuming a uniform discretization of $t$ with step $\var{t}$ and $f_l \equiv f(t_l)$. The term $\mathcal{L}{[f_l]}$ is given by:
\begin{multline}\label{eq:Lf}
    \mathcal{L}{[f_l]}=-\dfrac{1}{R\sin{\theta}} \bqty{ \pdv{ (f_l v_{\theta} \sin{\theta}) }{\theta} + \pdv{ (f_l v_{\varphi}) }{\varphi} } \\
        - \dfrac{\cos{\theta}}{R\sin{\theta}} \bqty{ v^2_{\varphi} \pqty{ \pdv{f}{v^{\hat{\theta}}} } - v_{\theta} \pqty{ \pdv{ (f v^{ \hat{\varphi}}) }{ v^{\hat{\varphi}}} } } 
        -\dfrac{1}{\tau} \bqty{ f_l - f^{\mathrm{eq}}_l }.
\end{multline}

\subsection{Summary} \label{sec:method:summary}

In this section, we summarize the key components required to implement the method described above. 
The first step is to select an appropriate quadrature order, based on the characteristics of the fluid flow under 
consideration. The $Q = 3$ quadrature may in general be sufficient for weakly compressible, isothermal flows,
whereas $Q = 4$ offers a more robust and stable approach allowing for the description of more complex flows. 
A choice of advection scheme must also be made from those presented in Section~\ref{sec:method:advection}.

The algorithm then proceeds using the multi-stage time-stepping procedure outlined in Section~\ref{sec:method:time}, 
where the central computational task is the evaluation of the $\mathcal{L}[f_l]$ operator. 
This process can be summarized as follows:

\begin{itemize}
    \item Angular derivatives in $\theta$ and $\varphi$ (from Eq.\eqref{eq:Lf}) are evaluated using the flux-based
     discretization in Eq.\eqref{eq:discrete_advection_sphere}, or alternatively using Eq.\eqref{eq:Komissarov_scheme} for
     the Komissarov-type advection scheme. The corresponding fluxes are defined in Eqs.\eqref{eq:fluxes_U1-3}--\eqref{eq:WENO5_sigma}.
    \item Velocity-space derivatives on the second line of Eq.\eqref{eq:Lf} are computed via matrix multiplication as shown
     in Eqs.\eqref{eq:df}, using the differentiation kernels given in Eqs.~\eqref{eq:Q3_kernels} and \eqref{eq:Q4_kernels} for quadrature orders $Q = 3$ and $Q = 4$, respectively.
    \item Macroscopic quantities such as density $n$ and velocity $\mathbf{u}$ are obtained from the distribution function moments
     using Eq.~\eqref{eq:f_moments_sum}.
    \item Equilibrium distributions $f^{\rm eq}_{\mathbf{k}}$ are then computed via Eqs.~\eqref{eq:feq_Q=3} and \eqref{eq:feq_Q=4}, depending on the chosen quadrature order.
\end{itemize}

\section{Numerical Results}\label{sec:numerics}
In this section, we present the numerical results obtained with the vielbein lattice Boltzmann 
scheme introduced in the previous section. 
We begin by examining the axisymmetric dynamics of sound and shear waves in Subsections~\ref{sec:numerics:sound} 
and \ref{sec:numerics:shear}, respectively, benchmarking our numerical solver against the analytical
solutions derived in Appendix~\ref{app:axi}. In Section~\ref{sec:numerics:riemann}, we analyze
the propagation of an axisymmetric shock wave originating from a discontinuous initial pressure profile, both centered on the pole (1D axisymmetric) and centered on the equator (2D simulation). Finally, we provide a qualitative analysis of the evolution of unsteady vortex configurations in a two-dimensional setup.

\subsection{Sound Wave Propagation}\label{sec:numerics:sound}

\begin{figure}[t]
    \centering
    \includegraphics[width=0.99\columnwidth]{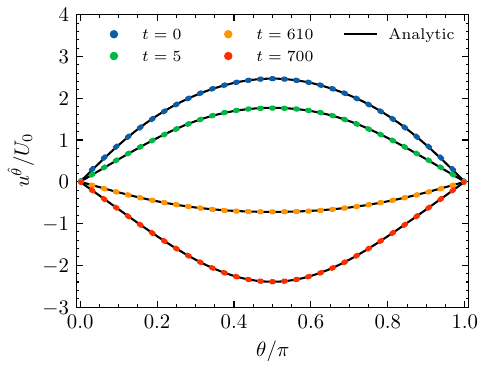}
    \caption{Full solution for the velocity profile of a sound wave with initial velocity $u^{\hat{\theta}}_0(\theta) = U_0 \theta(\pi - \theta)$. The velocity is normalized by the initial amplitude $U_0$, while the coordinate is shown as $\theta/\pi$.
    The numerical data (colored symbols) is shown to match the analytical solution (black lines) at every time $t$ (in lattice units) considered.
    }
    \label{fig:uth_sound_wave}
\end{figure}

In this section, we examine the capabilities of our numerical scheme to recover the evolution of a longitudinal (sound) wave, propagating axisymmetrically on the spherical surface. Aligning the symmetry axis of the flow configuration with the $z$ axis of our numerical grid, we define the longitudinal wave as a flow with a velocity profile $\vb{u}  = u^{\hat{\theta}} \vb{e}_{\theta}$ along the polar coordinate $\theta$, exhibiting a dependency on that coordinate alone: $u^{\hat{\theta}} \equiv u^{\hat{\theta}}(t,\theta)$. 
We are therefore interested in profiles of the type $u^{\hat{\theta}} \to \delta u^{\hat{\theta}}(\theta, t)$ and assume homogeneity along the $\varphi$ coordinate.

The problem described above is studied analytically with full details in Appendix~\ref{app:axi:sound}. Here, we only list the salient features of the setup and the solution. We consider small perturbations in the density $\rho$, pressure $P$, and velocity $u^{\hat{\theta}}$ over a background fluid at rest. The relevant fluid equations, shown in Eqs.~\eqref{eq:sound_eqs} and \eqref{eq:sound_diss_eqs}, can be merged by eliminating the pressure and density fluctuations in favor of the velocity fluctuations, leading to:
\begin{multline}\label{eq:eq_dissipative_sound}
    \pdv[2]{u^{\hat{\theta}}}{t} -  \frac{2 \nu }{R^2} \pdv{u^{\hat{\theta}}}{t} - \frac{c^2_{s}}{R^2} \pdv{\theta}\left[\frac{\partial_\theta(u^{\hat{\theta}} \sin\theta)}{\sin{\theta}}\right]\\
    + \frac{\nu + \nu_v}{R^2} \pdv{\theta} \left[ \frac{1}{\sin{\theta}} 
    \pdv{\theta}\left(\pdv{u^{\hat{\theta}}}{t} \sin\theta \right) \right] = 0,
\end{multline}
where $c_s = \partial P / \partial \rho$ is the speed of sound, while $\nu = \eta/\rho$ and $\nu_v = \zeta / \rho$ are the shear and volumetric kinematic viscosities, respectively.

We perform our benchmark test for an initial velocity and density of the type:
\begin{equation}\label{eq:sound_init}
u^{\hat{\theta}}_0(\theta) = U_0 (\pi - \theta) \theta, \quad 
\delta \rho_0(\theta) = 0,
\end{equation}
with $U_0$ as the initial amplitude. The initial state is even under the reflection $\theta \to \pi - \theta$. 
The above equation can be solved using a mode expansion, given in Eq.~\eqref{eq:sound_sol}. Our choice of initial profile selects only the even modes,
\begin{align}\label{eq:sound_modes_even}
 u^{\hat{\theta}}(t, \theta) &= \sum_{n = 1}^\infty \frac{U^e_n(t) F^e_n(\theta)}{\sin\theta},
\end{align}
where we retained only the terms proportional to the even angular functions, $F^e_n(\theta)$. The first few such functions can be evaluated using the explicit representation \eqref{eq:Jacobi_series} for the Jacobi polynomials in Eq.~\eqref{eq:sound_Fen}:
\begin{gather}
    F^e_1(\theta) = \frac{\sqrt{3}}{2} \sin^2 \theta, \quad 
    F^e_2(\theta) = \sqrt{\frac{21}{32}} \sin^2 \theta(1 - 5 \cos^2 \theta), \nonumber\\
    F^e_3(\theta) = \frac{\sqrt{165}}{16} \sin^2\theta (1 - 14 \cos^2 \theta + 21 \cos^4 \theta).
\end{gather}

\begin{figure*}[t]
    \includegraphics[width=\linewidth]{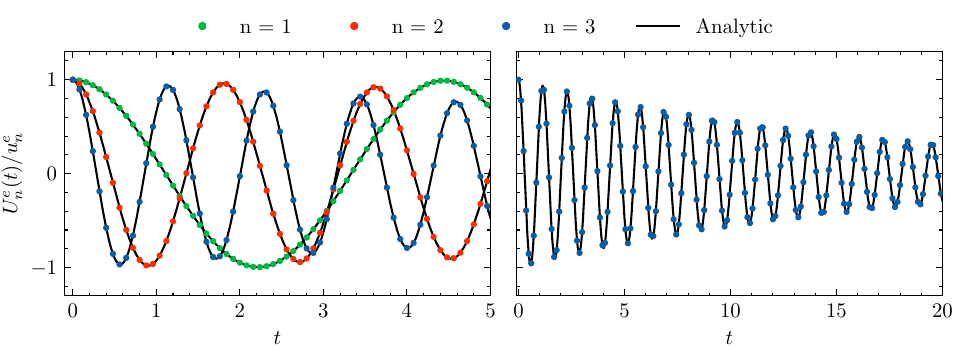}
    \caption{Evolution of different amplitudes for the sound wave solution. In the left panel, the numerical data (colored symbols) for the modes $n=1,2,3$ is plotted as a function of time, matching the analytical solution (black line) everywhere. In the right panel, the exponential decay of the fastest decaying mode considered ($n=3$) is made apparent by considering a larger time window. The solutions are normalized by the initial velocity amplitude $U_0$ and by the integration constant. 
    \label{fig:sound_ampl_evolution} 
    }
\end{figure*}

\begin{figure*}[htb]
    \includegraphics[width=0.99\textwidth]{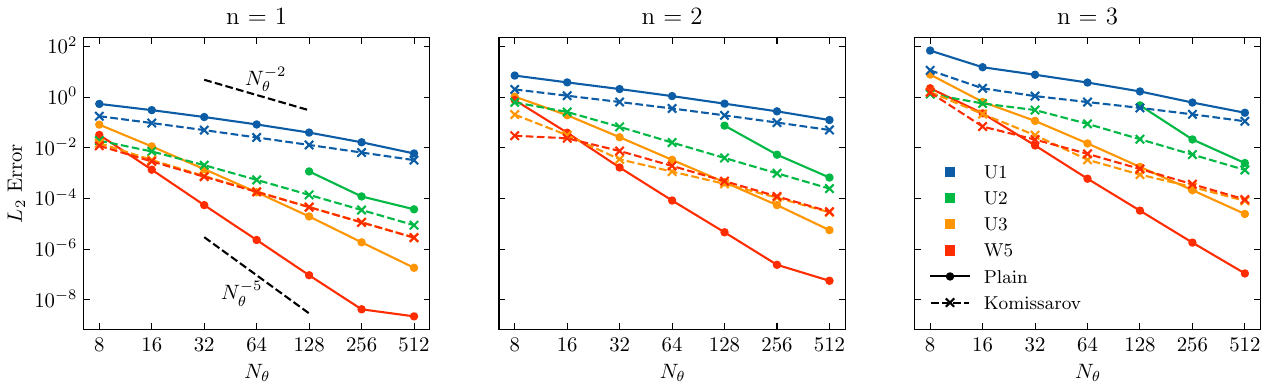}
    \caption{Convergence test for the sound wave solution. The L2 error evaluated between the analytical and numerical solution for (from left to right) the $n=1$, $n=2$ and $n=3$ modes, using different advection schemes, is plotted against the grid size $N_\theta$. U1, U2 and U3 indicate the first, second and third-order upwind schemes, while W5 is the \ac{weno5} scheme. Dotted lines refer to simulations in which we have employed the Komissarov scheme.
    }
    \label{fig:sound_convergence}
\end{figure*}

The time-dependent amplitudes $U^e_n(t)$ appearing in Eq.~\eqref{eq:sound_modes_even} are given by
\begin{equation}\label{eq:sound_amplitudes_aux}
 U^e_n(t) = u^e_n e^{-\gamma^e_n t} \cos[\omega^e_n(\nu) t + \varphi^e_n],
\end{equation}
where $\gamma^e_n$ and $\omega^e_n$ are given in Eq.~\eqref{eq:sound_damping_coeffs}.
In this section, we will consider the case when the shear and bulk viscosities are equal, $\nu = \nu_v$, such that the above coefficients reduce to:
\begin{equation}
 \gamma^e_n = \frac{\nu}{R^2} (\lambda_{e;n}^2 - 1), \quad
 \omega^2_{e;n}(\nu) = \omega^{2}_{e;n}(0) - \gamma_{e;n}^2,
\end{equation}
with $\lambda^2_{e;n} = 2n(2n-1)$ and $\omega^e_n = \lambda_n^e c_s / R$ is the acoustic angular frequency in the absence of dissipation.

The constants $u^e_n$ and $\varphi^e_n$ in Eq.~\eqref{eq:sound_amplitudes_aux} are fixed via Eqs.~\eqref{eq:sound_un}. Since $\partial \delta \rho_0 / \partial \theta = 0$, Eq.~\eqref{eq:sound_unsin} shows that 
\begin{equation}
 u_n^e \sin\varphi^e_n = \frac{\gamma_n^e}{\omega_n^e} u_n^e \cos\varphi^e_n,
\end{equation}
while $\mathfrak{u}_n = u^e_n \cos \varphi^e_n$ can be evaluated using Eq.~\eqref{eq:sound_uncos}, with the velocity profile in Eq.~\eqref{eq:sound_init}:
\begin{gather}
 \frac{\mathfrak{u}_1}{U_0} = \frac{\pi}{8 \sqrt{3}}(\pi^2 + 3), \quad 
 \frac{\mathfrak{u}_2}{U_0} = -\frac{\pi \sqrt{7}}{256\sqrt{6}}(4\pi^2 - 33), \nonumber
 \\
 \frac{\mathfrak{u}_3}{U_0} = \frac{\pi \sqrt{55}}{2048\sqrt{3}}(4\pi^2 - 37).
\end{gather}
With the above notation, $U^e_n(t)$ can be written as 
\begin{equation}\label{eq:sound_amplitudes}
 U^e_n(t) = \mathfrak{u}_n e^{-\gamma^e_n t} \left(\cos[\omega^e_n(\nu) t] - \frac{\gamma_n^e}{\omega_n^e} \sin[\omega^e_n(\nu) t]\right).
\end{equation}

We now discuss the validation of the above solution using our numerical solver.
Since the problem is axisymmetric, the solution is effectively 1D in space. Thus, we discretize the spatial grid using $N_\varphi= 1$ nodes along the azimuthal direction and various values for the number of nodes $N_\theta$ along the polar direction. In Fig.~\ref{fig:uth_sound_wave}, we compare the numerical (colored symbols) and analytical (solid line) solution for the velocity $u^{\hat{\theta}}(t,\theta)/U_0$ as a function of $\theta$, for selected time instances $t$. The simulation was run employing the D2Q16 stencil with the \ac{weno5} scheme for $N_\varphi=1$, $N_\theta=256$, $\var{t}=\num{2e-5}$, $\tau=\num{2e-5}$; the initial density was taken as $\rho=1$ everywhere, whereas the initial amplitude was set as $U_0=\num{1e-5}$ (everything is expressed in terms of lattice units (LU)). 
For all results presented in this work, the radius of the spherical surface was kept fixed at $R=1$. An excellent agreement is found between the numerical and analytical solutions.

A further check of our numerical code is shown in Fig.~\ref{fig:sound_ampl_evolution}, where we considered the time evolution of the amplitudes $U^e_n(t)$ for $n=\{1,2,3\}$. The simulation parameters used in this case were: $N_\varphi=1$, $N_\theta=512$, $\delta t = \num{1e-4}$, $\tau=\num{2e-3}$, initial density $\rho_0 = 1$ and velocity $U_0 = \num{1e-5}$. We increased $\tau$ with respect to the value considered in Fig.~\ref{fig:uth_sound_wave} to make the damping more apparent.
As can be seen, higher-order modes correspond to higher frequencies. Taking a wider time window, the exponentially decaying envelope typical of damped harmonic oscillators, shown for the $n = 3$ mode, is easily recognizable.
\begin{figure*}[!ht]
    \includegraphics[width=\linewidth]{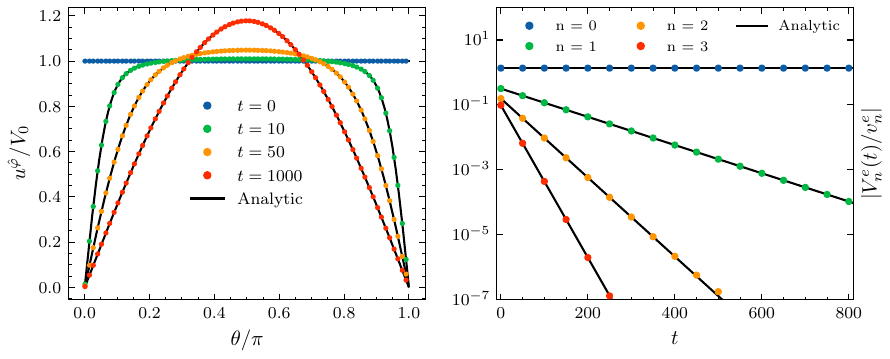}
    \caption{Comparison between the analytical and numerical solutions for the shear wave problem discussed in Sec.~\ref{sec:numerics:shear}. 
    (left panel) Velocity profiles for $u^{\hat{\varphi}}(t,\theta) / V_0$, normalized with respect to the initial 
    velocity amplitude $V_0 = 10^{-5}$, at selected time moments, shown as functions of the normalized polar coordinate $\theta/\pi$.
    (right panel) Normalized amplitudes $V_n^e(t) / V_0$ of the even modes $G_n^e(\theta)$, corresponding to $n=0,1,2,3$, 
    shown as functions of time. The numerical (colored symbols) and analytical (black lines) results agree down to $\sim \num{1e-7}$.
    }\label{fig:shear_solution}
\end{figure*}

In order to estimate the accuracy of our results, we take into account the $L_2$ relative error measured from the solutions' amplitudes and study the method's convergence for an increasing number of grid points along $\theta$. For each mode $n$, we evaluate the error as:
\begin{equation}\label{eq:l2_err}
    {\rm Err}_{L_2} = \sqrt{ \dfrac{ \int_{0}^{T} { \dd{t} \abs{ U_{e;n}^{\rm sim}(t) - U_{e;n}^{\rm th}(t) }^2 } } 
                           { \int_{0}^{T} { \dd{t} \abs{U_{e;n}^{\rm th}(t) }^2 } } }
    ,
\end{equation}
with $U_{e;n}^{\rm sim}$ and $U_{e;n}^{\rm th}$ being the amplitudes obtained numerically and analytically, respectively. The integration is performed over the total simulation time, $T = 20$.
Several numerical advection schemes were implemented and tested. In particular, we considered the
first-order (U1), second-order (U2) and third-order (U3) upwind schemes, together with the already mentioned \ac{weno5} scheme, presented in Sec.~\ref{sec:method:advection}. For the lower-order schemes, some advantage in the accuracy seemed to be visible when the fluxes entering the finite differences were written following the Komissarov scheme. In Fig.~\ref{fig:sound_convergence}, we show the relative error corresponding to the $n=\{1,2,3\}$ modes evaluated with several schemes. The convergence appears consistent with the orders of the schemes employed, reaching fifth order for \ac{weno5}. In the case of the Komissarov scheme, the convergence order is at most two, regardless of the scheme employed. The simulations were run with $N_\varphi=1$, $\var{t}=\num{2e-5}$, $\tau=\num{2e-5}$, initial density $\rho_0 = 1$ everywhere and initial amplitude $U_0=\num{1e-5}$, while the numbers of nodes along $\theta$ were taken as $N_\theta=[4,8,16,32,64,128,256,512]$. The same problem considered in a rotated coordinate system is presented in Fig.~\ref{fig:isotropy_test}, highlighting the isotropy of our scheme.

\subsection{Shear Wave Dissipation}\label{sec:numerics:shear}
\begin{figure}[htb]
    \includegraphics[width=0.99\columnwidth]{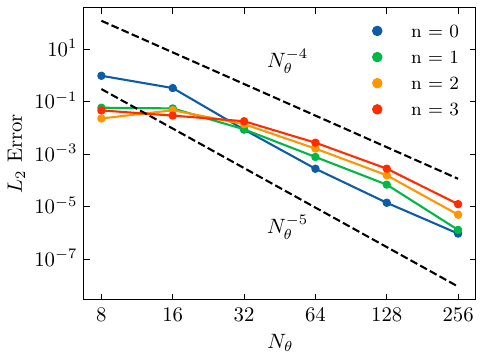}
    \caption{Convergence test for the shear wave solution. The L2 error evaluated between the analytical
             and numerical solution for the $n=0,1,2,3$ modes using the \ac{weno5} scheme 
             is plotted against the grid size $N_\theta$.
            }\label{fig:shear_convergence}
\end{figure}
In this section, we test the capabilities of our numerical scheme in the context of damping of axisymmetric shear waves on the sphere. The complete derivation of the analytical solution is presented in Appendix~\ref{app:axi:shear}. Here, we summarize the main features of this solution. As in Subsec.~\ref{sec:numerics:sound}, we take the vertical ($z$) axis of the grid to coincide with the symmetry axis of the flow. Therefore, the velocity profile is purely azimuthal, $\vb{u} = u^{\hat{\varphi}} \vb{e}_{\hat{\varphi}}$, while $u^{\hat{\varphi}} \equiv u^{\hat{\varphi}}(t, \theta)$ depends only on the polar coordinate $\theta$. The evolution equation for $u^{\hat{\varphi}}$, given in Eq.~\eqref{eq:shear_duphidt}, can be obtained by discarding terms that are quadratic in fluctuations in the axisymmetric Cauchy equation \eqref{eq:axi_NSE_uth},
and is given by:
\begin{equation}\label{eq:shear_equation}
 \frac{\partial}{\partial t} \left(\frac{u^{\hat{\varphi}}}{\sin\theta}\right) = \frac{\nu}{R^2 \sin^3\theta} \frac{\partial}{\partial \theta} \left[\sin^3 \theta \frac{\partial}{\partial \theta} \left(\frac{u^{\hat{\varphi}}}{\sin\theta}\right)\right].
\end{equation}

In this section, we consider the following initial velocity profile:
\begin{equation}
 u^{\hat{\varphi}}_0(\theta) = V_0 = {\rm const}.
 \label{eq:shear_u0}
\end{equation}
Since $u^{\hat{\varphi}}_0(\theta) = u^{\hat{\varphi}}_0(\pi - \theta)$ is even with respect to $\theta \rightarrow \pi - \theta$, the general solution \eqref{eq:shear_uphi_sol} can be written in terms of only the even functions $G_n^e(\theta)$ 
\begin{equation}\label{eq:shear_solution}
 u^{\hat{\varphi}}(t,\theta) = \sin\theta \sum_{n = 0}^\infty V^e_n(t) G^e_n(\theta), \quad 
 V_n^e(t) = v^e_n e^{-\kappa^e_n t},
\end{equation}
where $\kappa^e_n = 2n (2n + 3) \nu / R^2$ are the damping coefficients [see Eq.~\eqref{eq:shear_kappaen}]. The first few even angular functions $G^e_n(\theta)$ can be obtained by replacing the explicit series representation \eqref{eq:Jacobi_series} of the Jacobi polynomials in the formula \eqref{eq:shear_Gen},
\begin{align}
 G^e_0(\theta) &= \frac{\sqrt{3}}{2}, \quad 
 G^e_1(\theta) = \sqrt{\frac{21}{32}} (3 + 5 \cos 2\theta), \nonumber
 \\
 G^e_2(\theta) &= \frac{\sqrt{165}}{128} (15 + 28 \cos2\theta + 21 \cos 4\theta), \\
 G^e_3(\theta) &= \frac{\sqrt{105}}{2048} (350 + 675 \cos 2\theta + 594 \cos 4\theta + 429 \cos 6\theta) \nonumber.
\end{align}
Finally, the coefficients $v^e_n$ appearing in Eq.~\eqref{eq:shear_solution} can be obtained by substituting Eqs.~\eqref{eq:shear_u0} and \eqref{eq:shear_Gen} into Eq.~\eqref{eq:shear_veon}. The first few coefficients $v^e_n$ evaluate to:
\begin{gather}
 \frac{v^e_0}{V_0} = \frac{\pi\sqrt{3}}{4}, \quad 
 \frac{v^e_1}{V_0} = \frac{\pi \sqrt{21}}{32\sqrt{2}}, \nonumber
 \\
 \frac{v^e_2}{V_0} = \frac{\pi \sqrt{165}}{256}, \quad
 \frac{v^e_3}{V_0} = \frac{25\pi \sqrt{105}}{8192},
 \label{eq:shear_ampl}
\end{gather}
while the general term is
\begin{equation}
 \frac{v^e_n}{V_0} = \frac{\pi \sqrt{(2n+1)(4n+3)}}{4(n+1)^{3/2}} \left[\frac{1}{2^{2n}} 
 \binom{2n}{n} \right]^2.
\end{equation}

It is interesting to note that the $n=0$ term, corresponding to $G^e_0(\theta) = \sqrt{3}/2$, has a constant amplitude $V^e_0(t) = \pi\sqrt{3}/4$. Thus, it is expected that the corresponding mode exhibits no damping over time, as was similarly observed for the case of a shear wave in the torus geometry in Ref.~\cite{busuioc-jfm-2020}. The presence of this non-damped mode leads to the appearance of a nonzero asymptotic profile, $\lim_{t\to \infty} u^{\hat{\varphi}}(t) = \sin{\theta} G_0^e(\theta) V_0^e(t) = (3\pi / 8) \sin\theta$. The numeric solution obtained with our method is able to accurately match this analytic solution, as shown in the left panel of Fig.~\ref{fig:shear_solution}. The right panel of the same figure shows the first four amplitudes $V^e_n(t)$, $n = 0, \dots 3$, plotted as functions of time. The numerical data show an agreement with the analytical solution up to $\sim \num{1e-7}$, and our method correctly recovers the constant behavior of the $n=0$ mode. The simulation parameters used were:
$N_\varphi=1$, $N_\theta=512$, $\var{t}=\num{1e-4}$, $\tau=\num{1e-3}$, initial density $\rho=1$ everywhere and initial amplitude $V_0=\num{1e-5}$.
In Fig.~\ref{fig:shear_convergence} we show the L2-error for the first four modes as a function of the grid size, employing the \ac{weno5} scheme; as it can be seen, the convergence is approximately found to be between 4th and 5th orders.

\subsection{Riemann problem}\label{sec:numerics:riemann}

\begin{figure*}
    \includegraphics[width=0.99\textwidth]{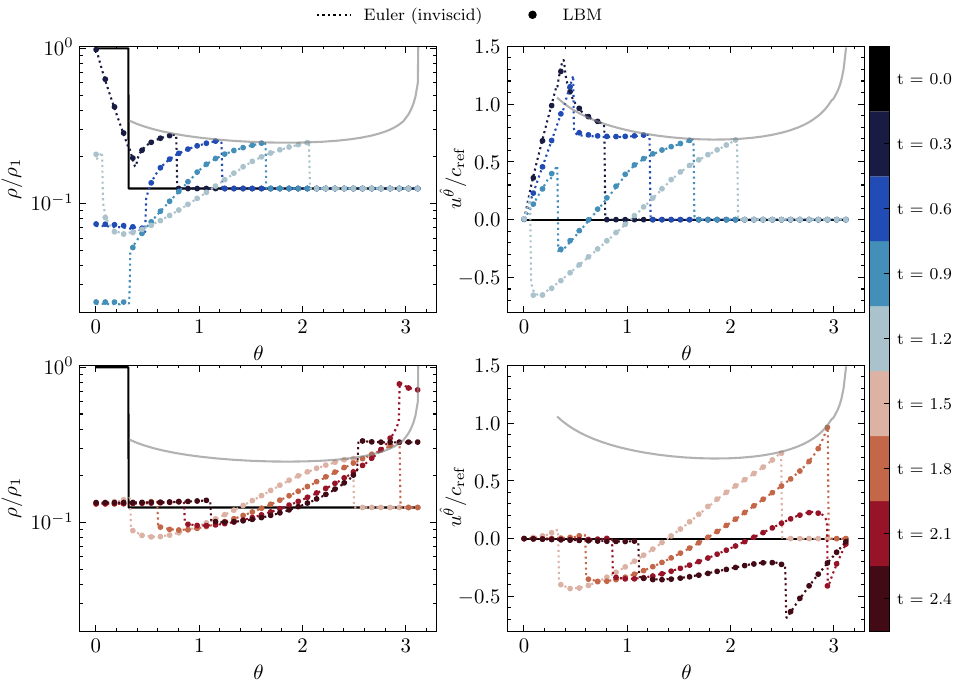}
    \caption{Density $\rho$ and polar velocity component $u^{\hat{\theta}}$ profiles, 
             normalized with respect to initial density $\rho_0$ and sound speed in the left domain, 
             shown at a few selected time instances: in the top row, we show the early stages up to the collapse of the reverse shock onto the north pole; in the bottom row, the late evolution stage, covering the arrival of the shock front at the south pole. The black thick lines show the initial conditions.
             Dotted lines show the numerical solution of the inviscid Euler equations (see Appendix~\ref{app:Euler}), and dots show the LBM results. Gray curves show Rankine-Hugoniot estimates for the shock front.}
    \label{fig:riemann}
\end{figure*}

In this section, we consider the spherical analogue of the classical Sod shock-tube problem. 
This test is used to assess the robustness of our numerical method in the presence of shock waves traveling at supersonic speed.

We take into consideration the following initial conditions:
\begin{align}
 \rho(\theta) = \begin{cases}
  \rho_1, & \theta < \theta_{\rm disc},\\
  \rho_2, & \theta > \theta_{\rm disc},
 \end{cases},
 \label{eq:riemann_ic}
\end{align}
with $\rho_1 =1 $,  $\rho_2 = 0.125$ and $\theta_{\rm disc}=\pi/10$.  At the inital time ($t = 0$) the fluid is at rest, with $u^\theta = u^\varphi = 0$.
The initial discontinuity in the density profile will give rise to a shock wave propagating towards the south pole,
with the discontinuity expected to remain sharp in the inviscid (perfect fluid) limit. 

Since no closed-form solution exists for shock wave propagation on a sphere, we employ two different approaches
to validate the accuracy of our LBM solver. First, we solve numerically the inviscid Euler equations on the sphere using 
a flux vector splitting method (details provided in Appendix~\ref{app:Euler}), yielding a high-resolution reference 
for the zero-viscosity limit.
Second, and as a novel contribution of this work, in what follows we derive a semi-analytical estimate of the 
shock front propagation by applying the Rankine-Hugoniot jump conditions to an axisymmetric discontinuity on the sphere.

We denote with $g(t,\theta)$ a generic physical quantity with a moving discontinuity at $\theta = \theta_d(t)$:
\begin{equation}
 g(t,\theta) = \Theta(\theta_d - \theta) g_N(t,\theta) + \Theta(\theta - \theta_d) g_S(t,\theta),
\end{equation}
where $\Theta$ is the Heaviside step function, and $g_{N/S}$ are the values north/south of the discontinuity. 
The time and angular derivatives of $g(t,\theta)$ read:
\begin{align}
 \dot{g} &= \Theta(\theta_d - \theta) \dot{g}_N + \Theta(\theta - \theta_d) \dot{g}_S 
    + \dot{\theta}_d \delta(\theta_d - \theta)(g_N - g_S), \nonumber\\
 g' &= \Theta(\theta_d - \theta) g'_N + \Theta(\theta - \theta_d) g'_S 
    - \delta(\theta_d - \theta)(g_N - g_S),
\end{align}
where $\delta(x)$ is the Dirac delta function.

Integrating the Euler equations \eqref{eq:Euler_cons} over an infinitesimal region around $\theta_d$ yields 
the Rankine-Hugoniot jump conditions:
\begin{subequations}\label{eq:Rankine-Hugoniot}
\begin{align}
 \rho_S (u_S - R \dot{\theta}_d) &= \rho_N(u_N - R \dot{\theta}_d), \\
 \rho_S (u_S^2 + c_s^2 - R \dot{\theta}_d u_S) &= \rho_N (u_N^2 + c_s^2 - R \dot{\theta}_d u_N),
\end{align}
\end{subequations}
where $R\dot{\theta}_d$ denotes the shock front velocity.
Solving these equations gives:
\begin{subequations}\label{eq:shock_gen}
\begin{equation}
 u_N - u_S = \pm c_s \left(\sqrt{\frac{\rho_N}{\rho_S}} - \sqrt{\frac{\rho_S}{\rho_N}}\right),
\end{equation}
with the sign chosen based on the shock propagation direction. The shock velocity is
\begin{equation}
 R \dot{\theta}_d = \frac{\rho_N u_N - \rho_S u_S}{\rho_N - \rho_S}.
\end{equation}
\end{subequations}

For the primary shock, where $u_S = 0$, $u_N > 0$, and $\rho_N > \rho_S$, this reduces to:
\begin{equation}
 R \dot{\theta}_d = \frac{\rho_N u_N}{\rho_N - \rho_S}, \quad 
 u_N = c_s \left(\sqrt{\frac{\rho_N}{\rho_S}} - \sqrt{\frac{\rho_S}{\rho_N}}\right).
 \label{eq:RH_primary}
\end{equation}

Given the shock velocity $R\dot{\theta}_d$ (which we extract from numerical data) and the known 
density $\rho_S = n_2$, the quantities north of the discontinuity can be computed as:
\begin{equation}
 u_N = \frac{(R\dot{\theta}_d)^2 - c_s^2}{R \dot{\theta}_d}, \quad 
 \rho_N = \rho_S \left(\frac{R \dot{\theta}_d}{c_s}\right)^2.
\end{equation}
These relations allow modeling the shock front dynamics, which we overlay in Fig.~\ref{fig:riemann} as 
solid gray lines for both velocity and density.

For LBM, we use a 1D spherical grid with resolution $N_\theta = 1024$, $N_\varphi = 1$, employing the \ac{weno5} 
advection scheme and a time step $\Delta t = 5 \times 10^{-6}$.

Figure~\ref{fig:riemann} shows the time evolution of the density and velocity, normalized 
with respect to the initial density $\rho_1$ and sound speed $c_s$. 
The top row corresponds to early times $t \in \{3, 6, 9, 12\}$, capturing the rarefaction wave collapse and formation of 
a reverse shock. The bottom row shows late-time snapshots $t \in \{15, 18, 21, 24\}$, including the shock front 
arrival at the south pole and its reflection.

The results of our LBM solver (marked with dots) are in perfect agreement with the numerical solution of the 
inviscid Euler equations (dotted lines), and both are also consistent with the modeling of the shock front 
described by the solid gray lines.
These results convincingly demonstrate that our method remains robust and accurate even in the presence 
of strong shocks and compressible effects.

In Appendix~\ref{app:rot}, we supplement our analysis by solving the shock tube problem on a fully two-dimensional grid 
considering a propagation direction not aligned with the meridional axes (see Fig.~\ref{fig:riemann_2D}).

\subsection{Vortices on the sphere}\label{sec:numerics:vortices}

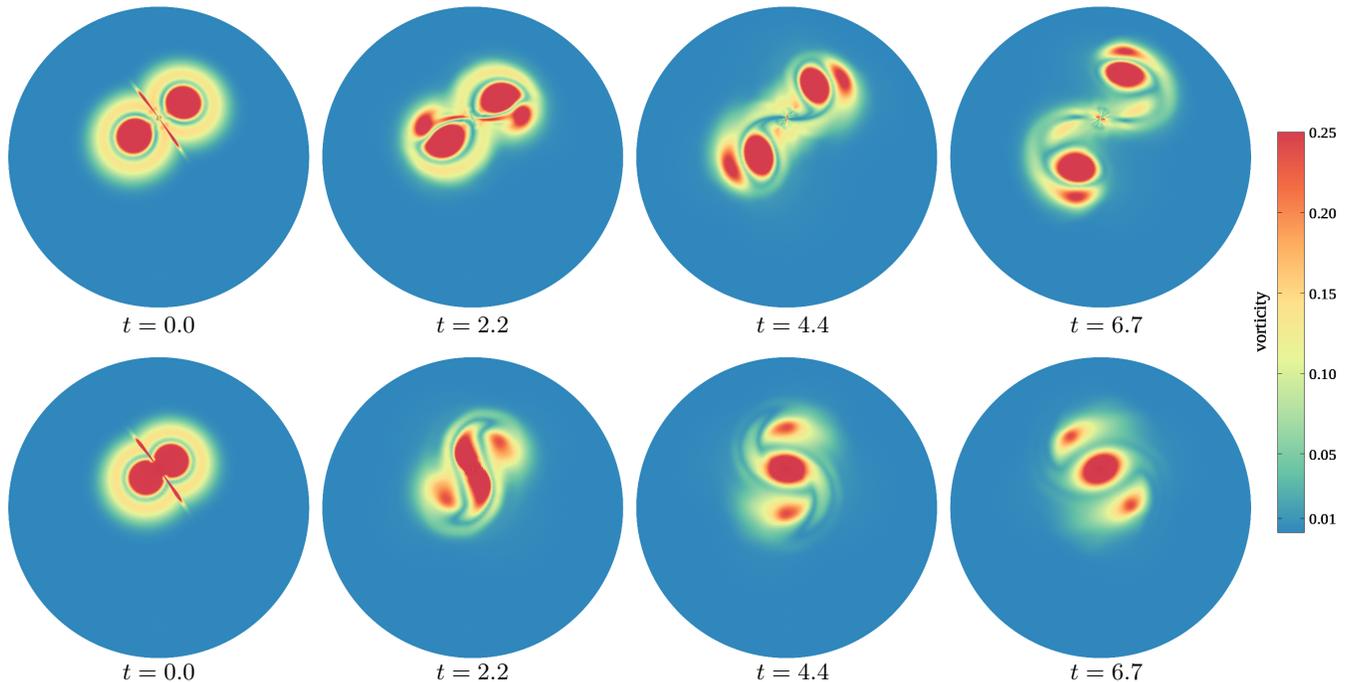
\begin{figure*}
    \centering
    \begin{overpic}[width=.99\textwidth]{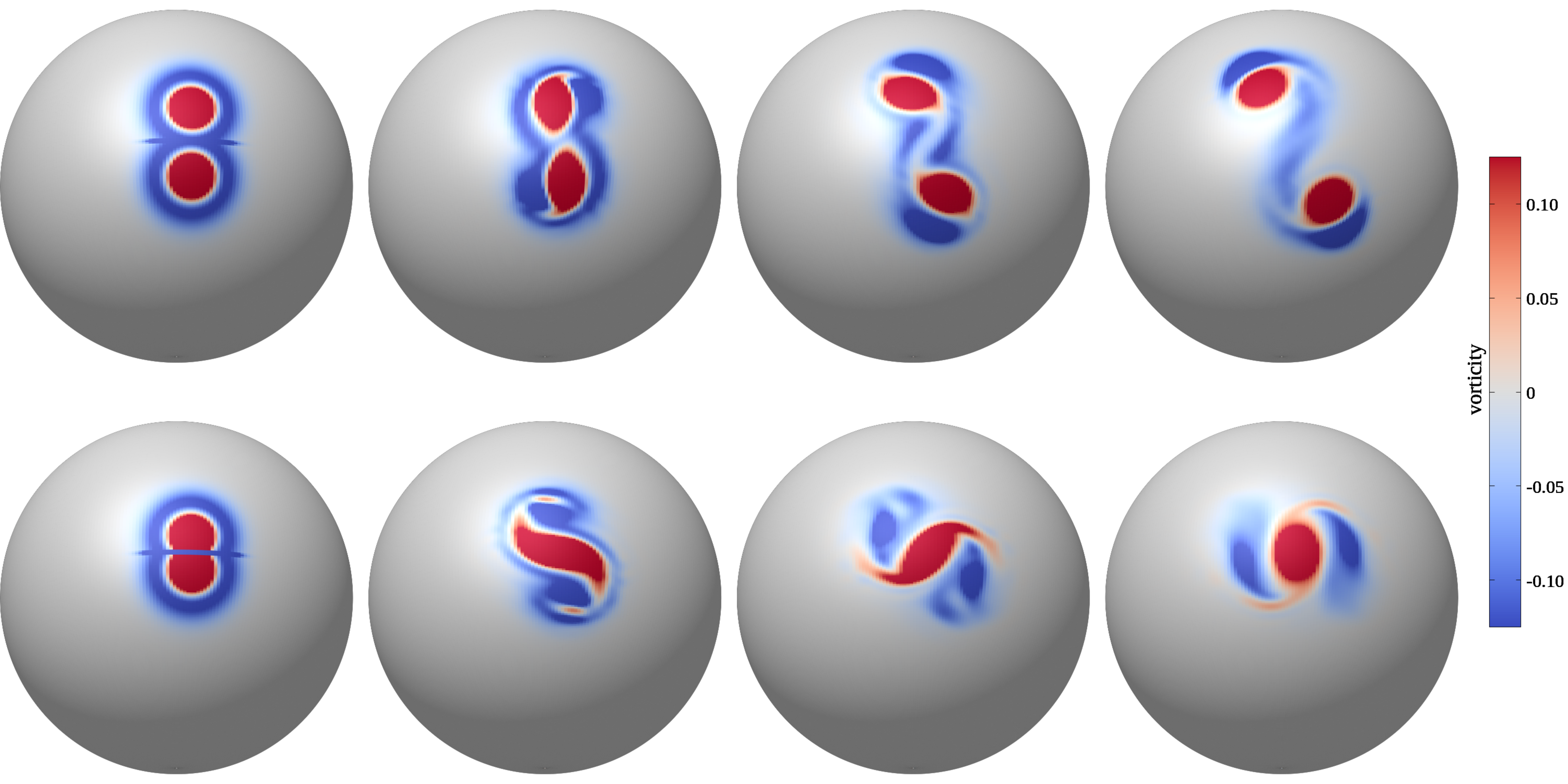}
     \put( 8.5,  25){$t = 0.0$}
     \put(32.0,  25){$t = 2.2$}
     \put(56.0,  25){$t = 4.4$}
     \put(79.5,  25){$t = 6.7$}
     \put( 8.5,  -1){$t = 0.0$}
     \put(32.0,  -1){$t = 2.2$}
     \put(56.0,  -1){$t = 4.4$}
     \put(79.5,  -1){$t = 6.7$}     
    \end{overpic}
    \caption{Temporal evolutions of the vorticity field for co-rotating vortex pairs. The initial position for the vortices' center is implicitly defined in Eq.~\ref{eq:vortex-ic}, with
    $c_1 = 0.8, c_2 = 1.2$ for the top row and
    $c_1 = 0.9, c_2 = 1.1$ on the bottom row. The corresponding computational times are shown below each snapshot.
            }\label{fig:vortex}
\end{figure*}

We conclude our numerical analysis with a fully two-dimensional test case, simulating the time evolution 
of co-rotating vortex pairs on the sphere, following the setup introduced in Ref.~\cite{yang-jcp-2020}. 
The dynamics of the system will depend on the initial distance at which the vortices are initialized.
We consider the following initial conditions for the velocity field:
\begin{equation}\label{eq:vortex-ic}
\begin{aligned}
    u &= 0,  \\
    v &= U_0 \begin{cases}
        -4(z - c_1) e^{0.3(1 - l_1^2)} & \text{if } z < 0, x > 0  \\
        -4(z - c_2) e^{0.3(1 - l_2^2)} & \text{if } z > 0, x > 0, \\
        0 & \text{otherwise,}
    \end{cases} \\
    w &= U_0 \begin{cases}
        4(y-1) e^{0.3(1 - l_1^2)} & \text{if } z < 0, x > 0  \\
        4(y-1) e^{0.3(1 - l_2^2)} & \text{if } z > 0, x > 0, \\
        0 & \text{otherwise,}
        \end{cases} 
\end{aligned}
\end{equation}
where $u, v, w$ are the Cartesian velocity components,
$l_1 = k \sqrt{\left(z-c_1\right)^2+(y-1)^2}$, $l_2 = k \sqrt{\left(z-c_2\right)^2+(y-1)^2}$, $k = 12.5$,
and $c_1$ and $c_2$ specify the position of the vortices' centers. 

We consider two representative cases: $(c_1, c_2) = (0.8, 1.2)$ and $(0.9, 1.1)$.
In both cases, we set the Reynolds number for the initial configuration as $\text{Re}= U_0 R / \nu = 3\times 10^{4}$, 
where the velocity amplitude was taken to be $U_0 = 0.1$, a constant density $\rho = 100$ and $\nu=\frac{10}{3}\times 10^{-6}$, 
in order to minimize compressibility effects (since the results reported in Ref.~\cite{yang-jcp-2020} have been obtained with a solver for the incompressible Navier-Stokes equations). 
A grid consisting of $(N_\theta, N_\varphi) = (320, 320)$ points was used, with a time step of $\delta t = 1 \times 10^{-6}$, and employing the WENO-5 numerical scheme.
The flow is visualized using the vorticity scalar field,
\begin{equation}
    \omega = \boldsymbol{\nabla} \cdot (\mathbf{u} \times \mathbf{e}_{\hat{r}}) 
    = 
    \frac{1}{r\sin\theta}\left( \frac{\partial(u\varphi \sin\theta)}{\partial \theta} - \frac{\partial u_\theta}{\partial \varphi} \right),
\end{equation}
where $\mathbf{e}_{\hat{r}}$ plays the role of the outwards-directed normal to the spherical surface \cite{rank-pof-2021}.

In Fig.~\ref{fig:vortex} we show snapshots of the vorticity at various times for both initial conditions,
which we find to be in qualitative agreement with the reference incompressible simulations of Ref.~\cite{yang-jcp-2020}.
In the first case (top row), the two vortices repel each other, remaining separated, and gradually drift away.
Conversely, when the initial distance of the two vortices is set below a threshold value, as in the second example (bottom row),
the vortices merge into a single structure. In this case we also observe the creation of two satellite vortices, 
which exhibit a dynamics similar to the one of the previous case, slowly drifting away from each other.

\section{Computational Performance}\label{sec:performances}
\begin{figure}[htb]
    \includegraphics[width=0.99\columnwidth]{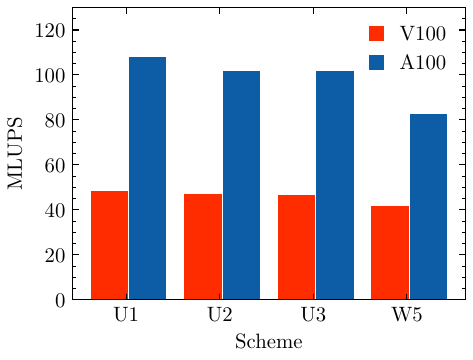}
    \caption{Computational performance in terms of millions of lattice updates per second (MLUPS) 
             for different advection schemes. We compare the results obtained 
             running on a single Nvidia V100 and a single Nvidia A100 GPU.
    }
    \label{fig:performance}
\end{figure}
In this section, we provide a brief overview of the numerical model's performance. 
We consider an implementation of vLBM using double-precision arithmetic and standard
optimization techniques~\cite{calore-pc-2016,calore-cc-2016} for GPU acceleration using
the OpenACC programming model.

Figure~\ref{fig:performance} presents the performance, measured in million lattice updates per second (MLUPS), 
for the different advection schemes discussed above. 
Performance data was obtained from the sound-wave benchmark on a fully 2D grid of size 
$1024 \times 1024$, running on a single Nvidia V100 and a single Nvidia A100 GPU.

Our results show that increasing the advection scheme from first-order upwind 
to third-order upwind incurs a negligible computational cost. 
However, the WENO-5 scheme introduces an overhead of approximately $\approx 20\%$ in the case of the A100 card. 
Nevertheless, for applications involving shock waves and supersonic flows, 
WENO-5 significantly enhances stability, making this overhead justifiable.

On average, the performance increases by about a factor of two going from the Nvidia V100 to the A100, consistent with the difference in peak computational capabilities between these architectures.

Compared to a standard on-grid LBM, the computational cost is approximately an order of magnitude higher. 
For reference, a similarly optimized isothermal D2Q17 model with a BGK collision operator achieves $\approx 1600$ 
MLUPS on an Nvidia A100 GPU. This additional cost arises from the Runge-Kutta time integration and off-grid advection scheme.

Notably, the Komissarov scheme has a negligible impact on the overall computational cost. 
Therefore, an optimal trade-off between performance and accuracy is achieved using
a third-order upwind scheme combined with the Komissarov scheme for enhanced stability. 
For simulations involving shock waves and supersonic flows, the WENO-5 scheme provides
the necessary stability, albeit at a moderate overhead.

\section{Conclusions}\label{sec:conclusions}

In this paper, we have developed a lattice Boltzmann model, based on the
vielbein formalism, for the simulation of fluid flows on spherical surfaces.
The vielbein fields encode all the geometric properties of the underlying
manifold, allowing the velocity space to be parametrized with respect to the
velocity components $(v^{\hat{\theta}}, v^{\hat{\varphi}})$ expressed with
respect to the vielbein.
This formulation trades the advantage of perfect streaming in traditional LBM
schemes for greater flexibility, enabling the description of fluid flows on
generic curved surfaces, as well as increased numerical stability. 
We leverage Gauss quadrature techniques to ensure the exact preservation of the
low-order moments of the particle distribution function, thereby retaining
one of the key properties of standard LBM formulations. This is in contrast
to finite difference discretizations of the Navier-Stokes equations or
discrete velocity methods, where such moment preservation is in general not guaranteed.

A key contribution of this work is the derivation of exact solutions to the Navier-Stokes equations for sound and shear wave dynamics on the sphere.  These analytical solutions, presented in Appendix~\ref{app:axi}, were used to validate and benchmark our numerical vLBM implementation. Additionally, they provide a valuable reference for future comparisons between different numerical solvers.

After successfully validating our model, we examined the propagation of an axisymmetric shock wave. As the shock travels from the north to the south pole, it is reflected upon reaching the latter. Simultaneously, a rarefaction wave depletes the north pole, leading to the formation of a reverse shock, both of which are captured in our simulations. 
To support the correctness of our results, we compared them with numerical solutions of the inviscid Euler equations and with a semi-analytic estimate of the shock front based on the Rankine-Hugoniot conditions. This benchmark highlights the capability of the solver to sustain the propagation of shock waves, also in the presence of sizeable compressibility effects.
We have also assessed the isotropy of the solver, comparing numerical results on an axisymmetric grid and a $90^{\circ}$-tilted configuration, finding excellent agreement.

As a final, more qualitative test, we examined the evolution of two vortices initialized on the spherical surface, following the setup suggested by Yang et al.~\cite{yang-jcp-2020}. We found a good qualitative agreement between our results and those reported therein.

Finally, we have reported a quick overview of the computational performances achieved by implementing the solver targeting modern GPU architectures. Our analysis suggests that an optimal trade-off between performance and accuracy is obtained using a third-order upwind scheme for advection. On the other hand, in the presence of flows subject to shock waves and supersonic speeds, the WENO-5 scheme may provide the necessary stability at a moderate overhead.

In future work, we plan to address key challenges in studying turbulence on
manifolds. To enable DNS at high Reynolds numbers, we will develop a 
multi-GPU implementation to efficiently handle high-resolution grids 
and large-scale simulations. Additionally, the current
explicit method imposes a time-step restriction due to the
Courant-Friedrichs-Lewy (CFL) condition, as well as the $\delta t < \tau$ limitation typical for explicit schemes, due to the stiffness of the collision term. To overcome this, 
we will incorporate implicit-explicit (IMEX) methods~\cite{pareschi-jsc-2010,wang-ijmpc-2007} or semi-Lagrangian implicit
approaches~\cite{chen-pre-2020} to enable coarser time steps, even for
high-order advection schemes. 
This will pave the way for large-scale simulations of turbulent flows on spherical surfaces. 
We also plan to explore the method’s performance beyond the hydrodynamic regime, leveraging its
kinetic foundation.

\section*{Data Availability}
All data are available upon reasonable request to the authors. 
A simplified implementation of the algorithm, alongside scripts and datasets to reproduce some of the results presented in this work, 
can be found in the following GitHub repository: \href{https://github.com/agabbana/vielbein_lbm}{https://github.com/agabbana/vielbein\_lbm}

\begin{acknowledgments}
We would like to acknowledge Alexander Wagner and P. Aasha for insightful discussions. We are also grateful to Andrei Mohu\cb{t} for doing the preliminary analysis for flows on a sphere as part of his dissertation project at the West University of Timi\cb{s}oara.
A.G. gratefully acknowledges the support of the U.S. Department of Energy through 
the LANL/LDRD Program under project number 20240740PRD1 
and the Center for Non-Linear Studies for this work.
VEA gratefully acknowledges LANL for kindly supporting a visit during which part of this work was completed.
E.B. acknowledges the support from the European Union’s HORIZON MSCA Doctoral Networks programme, under Grant Agreement No. 101072344, project AQTIVATE (Advanced computing, QuanTum algorIthms and data-driVen Approaches for science, Technology and Engineering).
\end{acknowledgments}

\appendix

\section{Velocity quadrature}\label{app:quad}

The discrete velocity set $\mathbf{v}_{\mathbf{k}} = (v^{\hat{\theta}}_{k_\theta}, v^{\hat{\varphi}}_{k_\varphi})$ is indexed using $\vb{k} = (k_\theta, k_\varphi)$, with $1 \le k_\theta, k_\varphi \le Q$. The corresponding weights $w_{\vb{k}} = w_{k_\theta} w_{k_\varphi}$ are computed using Eq.~\eqref{eq:GH_weights}.

Two popular quadratures employed in lattice Boltzmann modeling, namely the third-order, D2Q9 ($Q = 3$) and fourth-order, D2Q16 ($Q = 4$) quadratures, are discussed in Sections~\ref{app:quad:Q3} and \ref{app:quad:Q4}, respectively. We address the accuracy of these quadrature models by employing a Chapman-Enskog analysis in Subsec.~\ref{app:quad:CE}.

\subsection{Third-order quadrature (\texorpdfstring{$Q = 3$}{Q=3})}\label{app:quad:Q3}

\begin{figure*}[ht]
    \centering
    \includegraphics[width=0.99\columnwidth]{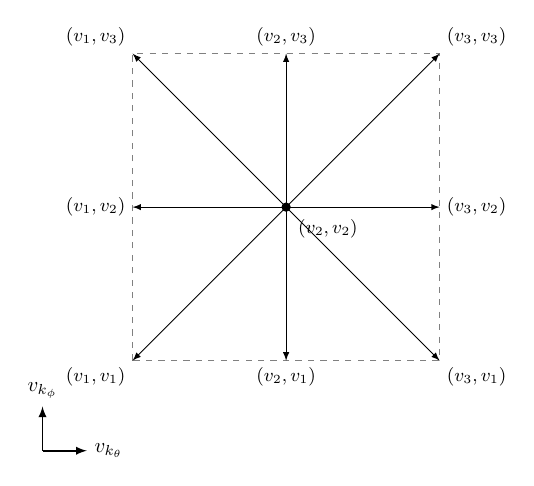}
    \includegraphics[width=0.99\columnwidth]{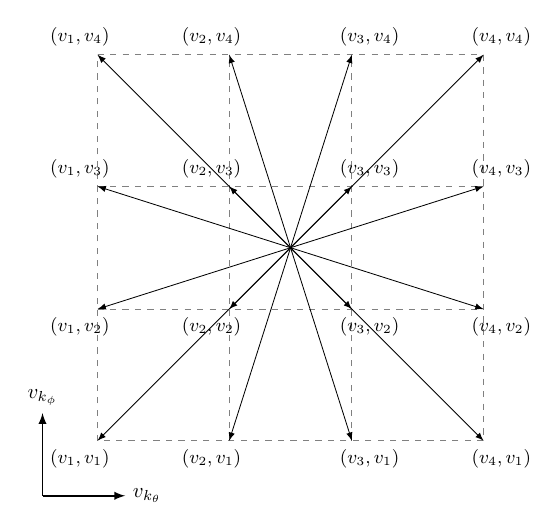}
    \caption{
    The left (right) panel shows the velocity set for the quadrature of order $Q=3$ ($Q=4$). Each arrow represents one of the velocity vectors $\vb{v}_k$ as listed in Tables~\ref{tab:velocity_set_Q=3}--\ref{tab:velocity_set_Q=4}. An additional circle at the center of the D2Q9 stencil denotes the inclusion of the null (zero) velocity vector.
For $Q = 3$, the velocity components are $(v_1, v_2, v_3) = (-\sqrt{3}, 0, \sqrt{3})$, corresponding to the roots of the third-order Hermite polynomial $H_3(x) = x^3 - 3x$.
For $Q = 4$, the velocity components are $v_4 = -v_1 = \sqrt{3 + \sqrt{6}}$ and $v_3 = -v_2 = \sqrt{3 - \sqrt{6}}$, which are the roots of the fourth-order Hermite polynomial $H_4(x) = x^4 - 6x^2 + 3$.
    }
    \label{fig:stencils}
\end{figure*}

The quadrature of order $Q=3$ corresponds to the velocity set shown in the left panel of Fig.~\ref{fig:stencils}. The corresponding weights $w_{\vb{k}} = w_{k_\theta} w_{k_\varphi}$ can be evaluated via \eqref{eq:GH_weights}, taking $H_4(x)=x^4 -6x^2 +3$ for the denominator. The nine resulting velocity vectors and their corresponding quadrature weights are summarized in \autoref{tab:velocity_set_Q=3}. 

\begin{table}
    \centering
    \begin{tabular}{lccc}
    \hline \hline
    $q$ & $v_{k_\theta}$ & $v_{k_\varphi}$ & $w_{\vb{k}}$ \\ [2pt]
    \hline 
    $         1$ & $               0$ & $                0$ & $4 / 9 $ \\ [2pt]
    $2 \ldots 5$ & $(\pm \sqrt{3},0)$ & $(0, \pm \sqrt{3})$ & $1 / 9 $ \\ [2pt]
    $6 \ldots 9$ & $ \pm \sqrt{3}   $ & $     \pm \sqrt{3}$ & $1 / 36$ \\ [2pt]
    \hline \hline
    \end{tabular}
    \caption{Components of the velocity vectors $\vb{v}_k$ and their weights $w_{\vb{k}}$ in the two-dimensional model of order $Q=3$, following the standard presentation conventions used in the literature \cite{kruger-book-2017}).}
    \label{tab:velocity_set_Q=3}
\end{table}

The equilibrium distribution function can be computed using $Q = 3$ in Eq.~\eqref{eq:feq_discrete}:
\begin{equation}\label{eq:feq_Q=3}
    f^{Q=3,\mathrm{eq}}_{\vb{k}}= n w_{\vb{k}} \bqty{1 + \vb{v_k} \vdot \vb{u} + \dfrac{1}{2} \pqty{ (\vb{v_k} \vdot \vb{u})^2 - u^2}}.
\end{equation}

The kernel matrices $\mathcal{K}^H_{k,k'}$ and $\widetilde{\mathcal{K}}^H_{k,k'}$ required to evaluate the velocity gradients can be obtained by setting $Q = 3$ in Eqs.~\eqref{eq:KH} and \eqref{eq:KtildeH}:
\begin{subequations}\label{eq:Q3_kernels}
\begin{align}
    \mathcal{K}_{k, k'}^H &=
    \pmqty{
    \frac{\sqrt{3}}{2} & \frac{1}{2 \sqrt{3}} & -\frac{1}{2 \sqrt{3}} \\
    -\frac{2}{\sqrt{3}} & 0 & \frac{2}{\sqrt{3}} \\
    \frac{1}{2 \sqrt{3}} & -\frac{1}{2 \sqrt{3}} & -\frac{\sqrt{3}}{2}
    }, \label{eq:Q3_Kk} \\
    \nonumber \\
    \widetilde{\mathcal{K}}_{k, k^{\prime}}^H &= \pmqty{
    -\frac{3}{2} & 0 & -\frac{1}{2} \\
    2 & 0 & 2 \\
    -\frac{1}{2} & 0 & -\frac{3}{2}
    } .
    \label{eq:Q3_KKt}
\end{align}
\end{subequations}

\subsection{Fourth-order quadrature (\texorpdfstring{$Q=4$}{Q=4})}\label{app:quad:Q4}
\begin{table}[ht]
    \centering
    \begin{tabular}{lccc}
    \hline \hline
    $q$ & $v_{k_\theta}$ & $v_{k_\varphi}$ & $w_{\vb{k}}$ \\ [2pt]
    \hline
    $1  \ldots 4 $ & $\pm \sqrt{3-\sqrt{6}}$ & $\pm \sqrt{3-\sqrt{6}}$ & $(5+2 \sqrt{6}) / 48$ \\ [2pt]
    $5  \ldots 8 $ & $\pm \sqrt{3+\sqrt{6}}$ & $\pm \sqrt{3-\sqrt{6}}$ & $1 / 48$              \\ [2pt]
    $9  \ldots 12$ & $\pm \sqrt{3-\sqrt{6}}$ & $\pm \sqrt{3+\sqrt{6}}$ & $1 / 48$              \\ [2pt]
    $13 \ldots 16$ & $\pm \sqrt{3+\sqrt{6}}$ & $\pm \sqrt{3+\sqrt{6}}$ & $(5-2 \sqrt{6}) / 48$ \\ [2pt]
    \hline \hline
    \end{tabular}
    \caption{Projections of the velocity vectors $\vb{v}_k$ and their weights $w_{\vb{k}}
    $ in the two-dimensional model of order $Q=4$ \cite{shan-jofm-2006,sofonea-pre-2018}, also referred to as the D2Q16 model.}
    \label{tab:velocity_set_Q=4}
\end{table}
The $4 \times 4= 16$ velocity vectors of the fourth-order, D2Q16 velocity set are displayed in the right panel of Fig.~\ref{fig:stencils} and are summarized, together with their corresponding weights, in Table~\ref{tab:velocity_set_Q=4}. The weights $w_{\vb{k}} = w_{k_\theta} w_{k_\varphi}$ are evaluated using Eq.~\eqref{eq:GH_weights}, with $H_{Q+1} = H_5(x)=x^5 -10x^3 + 15x$ in the denominator.

The expression of the equilibrium distribution function for the $Q=4$ model is:
\begin{multline}\label{eq:feq_Q=4}
    f^{Q=4,\mathrm{eq}}_{\vb{k}}= n w_{\vb{k}} \bigg[1 + \vb{v_k} \vdot \vb{u} 
    + \dfrac{1}{2} \pqty{ (\vb{v_k} \vdot \vb{u})^2 - u^2} \\
    + \dfrac{1}{6} \vb{v_k} \vdot \vb{u} \pqty{ (\vb{v_k} \vdot \vb{u})^2 - 3u^2} \bigg].
\end{multline}
The kernel matrices required to compute the velocity gradients, defined in Eqs.~\eqref{eq:KH} and \eqref{eq:KtildeH}, evaluate to
\begin{widetext}
\begin{subequations}\label{eq:Q4_kernels}
    \begin{align}
    \mathcal{K}_{k, k'}^H &= \pmqty{
    \frac{1}{2} \sqrt{3+\sqrt{6}} & \frac{\sqrt{3+\sqrt{3}}}{2(3+\sqrt{6})} & -\frac{\sqrt{3-\sqrt{3}}}{2(3+\sqrt{6})} & \frac{\sqrt{3-\sqrt{6}}}{2 \sqrt{3}} \\
    -\sqrt{\frac{5+2 \sqrt{6}}{2(3-\sqrt{3})}} & \frac{1}{2} \sqrt{3-\sqrt{6}} & \frac{1}{2} \sqrt{1+\sqrt{\frac{2}{3}}} & -\sqrt{\frac{5+2 \sqrt{6}}{2(3+\sqrt{3})}} \\
    \sqrt{\frac{5+2 \sqrt{6}}{2(3+\sqrt{3})}} & -\frac{1}{2} \sqrt{1+\sqrt{\frac{2}{3}}} & -\frac{1}{2} \sqrt{3-\sqrt{6}} & \sqrt{\frac{5+2 \sqrt{6}}{2(3-\sqrt{3})}} \\
    -\frac{\sqrt{3-\sqrt{6}}}{2 \sqrt{3}} & \frac{\sqrt{3-\sqrt{3}}}{2(3+\sqrt{6})} & -\frac{\sqrt{3+\sqrt{3}}}{2(3+\sqrt{6})} & -\frac{1}{2} \sqrt{3+\sqrt{6}}
    }, \label{eq:Q4_Kk}\\
    \nonumber \\
    \widetilde{\mathcal{K}}_{k, k'}^H &= \pmqty{
    -\frac{3+\sqrt{6}}{2} & \frac{2-5 \sqrt{2}+\sqrt{6(9-4 \sqrt{2})}}{4} & \frac{2+5 \sqrt{2}-\sqrt{6(9+4 \sqrt{2})}}{4} & \frac{1}{2} \\
    \frac{2+5 \sqrt{2}+4 \sqrt{3}+\sqrt{6}}{4} & -\frac{3-\sqrt{6}}{2} & \frac{1}{2} & \frac{2-5 \sqrt{2}-4 \sqrt{3}+\sqrt{6}}{4} \\
    \frac{2-5 \sqrt{2}-4 \sqrt{3}+\sqrt{6}}{4} & \frac{1}{2} & -\frac{3-\sqrt{6}}{2} & \frac{2+5 \sqrt{2}+4 \sqrt{3}+\sqrt{6}}{4} \\
    \frac{1}{2} & \frac{2+5 \sqrt{2}-\sqrt{6(9+4 \sqrt{2})}}{4} & \frac{2-5 \sqrt{2}+\sqrt{6(9-4 \sqrt{2})}}{4} & -\frac{3+\sqrt{6}}{2}
    }.
    \label{eq:Q4_KKt}
    \end{align}
\end{subequations}
\end{widetext}

\subsection{Conservation equations} \label{app:quad:cons}

The recovery of the mass and momentum conservation equations, Eqs.~\eqref{eq:continuity_covariant} and \eqref{eq:NSE_covariant}, requires the integration of Eq.~\eqref{eq:Boltzmann_eq_covariant} multiplied by $1$ and $v^{\hat{a}}$, over the velocity space. The discretization of the velocity space must ensure the exact recovery of the following integrals of the distribution function:
\begin{gather}
 \int d\mathbf{v}\, f = n, \quad 
 \int d\mathbf{v} \, f \mathbf{v} = n \mathbf{u}, \nonumber\\ 
 m\int d\mathbf{v}\, f v^{\hat{a}} v^{\hat{b}} = \rho u^{\hat{a}} u^{\hat{b}} - \tau^{\hat{a} \hat{b}}.
 \label{eq:quad_cons_moments}
\end{gather}
The structure of $\tau^{\hat{a}\hat{b}}$ and its connection to the constitutive relation for the Newtonian fluid is addressed in Subsec.~\ref{app:quad:CE}. In both the $Q = 3$ and $Q = 4$ quadrature models, the projection of $f$ onto the Hermite space, together with the Gauss integration rules, ensures the preservation of the integrals appearing above.

Within a model of order $Q$, the continuous distribution function $f(\vb{v})$ is replaced by the discrete sets $f_{\vb{k}}$ via Eq.~\eqref{eq:f_disc}, which allow for a Hermite series representation as given in Eq.~\eqref{eq:f_series}, truncated at $\ell_\theta = \ell_\varphi = Q - 1$:
\begin{equation}
 f(\vb{v}) \to f_{\vb{k}} = w_{\mathbf{k}} \sum_{\ell_\theta, \ell_\varphi = 0}^{Q - 1} \frac{\mathcal{F}_{\ell_\theta, \ell_\varphi}}{\ell_\theta! \ell_\varphi!} H_{\ell_\theta}(v^{\hat{\theta}}_{k_\theta}) H_{\ell_\varphi}(v^{\hat{\varphi}}_{k_\varphi}),
\end{equation}
which guarantees the recovery of the moments of the following type:
\begin{equation}
 \int d\mathbf{v}\, f(\mathbf{v}) v_{\hat{\theta}}^{\ell_\theta} v_{\hat{\varphi}}^{\ell_\varphi} = \sum_{\mathbf{k}} f_{\mathbf{k}} v_{k_\theta}^{\ell_\theta} v_{k_\varphi}^{\ell_\varphi},
\end{equation}
for any $0 \le \ell_\theta, \ell_\varphi < Q$. Eqs.~\eqref{eq:quad_cons_moments} require moments with $0 \le \ell_\theta, \ell_\varphi \le 2$, hence our $Q = 3$ and $Q = 4$ quadratures are sufficient.

Moreover, the following integrals of the velocity gradients of $f$ must be recovered:
\begin{align}
 \int d\mathbf{v} \frac{\partial(v^{\hat{b}} v^{\hat{c}} f)}{\partial v^{\hat{a}}} &= 0, \label{eq:quad_cons_dfdv_moments}\\
 \int d\mathbf{v} \frac{\partial(v^{\hat{b}} v^{\hat{c}} f)}{\partial v^{\hat{a}}} v^{\hat{d}} &= -\delta^{\hat{d}}_{\hat{a}} \int d\mathbf{v} \, f v^{\hat{b}} v^{\hat{c}}. 
 \label{eq:quad_cons_vdfdv_moments}
\end{align}
Specifically, in spherical coordinates, the following integrals must be ensured:
\begin{subequations}\label{eq:quad_cons_df_moments}
\begin{align}
 \int d\mathbf{v} \frac{\partial f}{\partial v^{\hat{\theta}}} v_{\hat{\varphi}}^2 &= \int d\mathbf{v} \frac{\partial f}{\partial v^{\hat{\theta}}} v_{\hat{\varphi}}^3 = 0, \\
 \int d\mathbf{v} \frac{\partial f}{\partial v^{\hat{\theta}}} v_{\hat{\varphi}}^2 v^{\hat{\theta}} &= -\int d\mathbf{v}\, f v_{\hat{\varphi}}^2, \\
 \int d\mathbf{v} \frac{\partial (fv^{\hat{\varphi}})}{\partial v^{\hat{\varphi}}} v^{\hat{\theta}} &= 
 \int d\mathbf{v} \frac{\partial (fv^{\hat{\varphi}})}{\partial v^{\hat{\varphi}}} v^2_{\hat{\theta}} = 0, \\
 \int d\mathbf{v} \frac{\partial (fv^{\hat{\varphi}})}{\partial v^{\hat{\varphi}}} v^{\hat{\theta}} v^{\hat{\varphi}} &= -\int d\mathbf{v}\, f v^{\hat{\varphi}} v^{\hat{\theta}}.
\end{align}
\end{subequations}
After velocity discretization, the velocity gradient terms are replaced via
\begin{subequations}
\begin{align}
    \pqty{ \pdv{f}{v^{\hat{\theta}}}}_{k_\theta,k_\varphi}   &= \sum_{k_\theta'=1}^{Q} \mathcal{K}^H_{k_\theta,k'_\theta} f_{k'_\theta,k_\varphi} \\
    \pqty{ \pdv{(f v^{\hat{\varphi}})}{v^{\hat{\varphi}}} }_{k_\theta,k_\varphi} &= \sum_{k'_\varphi = 1}^{Q} \widetilde{\mathcal{K}}^H_{k_\varphi,k'_\varphi} f_{k_\theta,k'_\varphi}.
\end{align}
\end{subequations}
It is straightforward to check that the following relations are satisfied by the kernels $\mathcal{K}^H_{k_\theta,k_\theta'}$ and $\widetilde{\mathcal{K}}^H_{k_\varphi,k_\varphi'}$ in both the $Q = 3$ \eqref{eq:Q3_KKt} and $Q = 4$ \eqref{eq:Q4_KKt} discretizations:
\begin{gather}
 \sum_{k_\theta = 1}^Q \mathcal{K}^H_{k_\theta, k'_\theta} = \sum_{k_\varphi = 1}^Q \widetilde{\mathcal{K}}^H_{k_\varphi, k'_\varphi} = 0, \nonumber\\
 \sum_{k_\theta = 1}^Q \mathcal{K}^H_{k_\theta, k'_\theta} v_{k_\theta} = -1, \ 
 \sum_{k_\varphi = 1}^Q \widetilde{\mathcal{K}}^H_{k_\varphi, k'_\varphi} v_{k_\varphi} = -v_{k_\varphi'}.
\end{gather}
Therefore, Eqs.~\eqref{eq:quad_cons_df_moments} are recovered by both the $Q = 3$ and $Q = 4$ quadratures considered in this paper.

\subsection{Chapman-Enskog analysis}\label{app:quad:CE}

In subsections~\ref{app:quad:Q3} and \ref{app:quad:Q4}, we presented the quadratures of orders $Q = 3$ and $Q = 4$. We discussed the recovery of the mass and momentum conservation equations within these discrete models in Sec.~\ref{app:quad:cons}. We now discuss the recovery of the Navier-Stokes equations, for which we consider the Chapman-Enskog expansion, following Ref.~\cite{ambrus-pre-2019}.

The constraint that $\psi \in \{1, \mathbf{v}\}$ represents collision invariants of the BGK model implies that the deviation $\delta f = f - f^{\rm eq}$ of the distribution function from equilibrium satisfies:
\begin{equation}
 \int d\mathbf{v}\, \delta f = 
 \int d\mathbf{v}\, \delta f \, v^{\hat{a}} = 0.
\end{equation}
Considering that the fluid is close to equilibrium, $\delta f$ can be taken as a small quantity, of the order of the relaxation time $\tau$. From Eq.~\eqref{eq:Boltzmann_eq_covariant}, $\delta f$ can be obtained to leading order as:
\begin{multline}
 \delta f \simeq -\tau \Bigg[ 
 \frac{\partial f^{\rm eq}}{\partial t} + 
 \frac{1}{\sqrt{g}} \frac{\partial}{\partial q^b} 
 \left(v^{\hat{a}} e_{\hat{a}}^ b f^{\rm eq} \sqrt{g}\right) \\
 - \Gamma^{\hat{a}}_{\hat{b}\hat{c}} \frac{\partial(v^{\hat{b}} v^{\hat{c}} f^{\rm eq})}{\partial v^{\hat{a}}} \Bigg],
 \label{eq:CE_df_gen}
\end{multline}
In the case of the spherical geometry, the vielbein vectors $e^b_{\hat{a}}$ and the connection coefficients $\Gamma^{\hat{a}}_{\hat{b}\hat{c}}$ are given in Eqs.~\eqref{eq:vielbein_frame_sphere} and \eqref{eq:connection_coefs_sphere}, respectively. Employing these expressions in Eq.~\eqref{eq:CE_df_gen} allows $\delta f$ to be obtained as
\begin{multline}
 \delta f \simeq -\tau \Bigg[ 
 \frac{\partial f^{\rm eq}}{\partial t} + 
 \frac{v^{\hat{\theta}} \partial_\theta(f^{\rm eq} \sin\theta)}{R\sin\theta} + 
 \frac{v^{\hat{\varphi}} \partial_\varphi f^{\rm eq}}{R\sin\theta} \\
 + \frac{\cos\theta}{R \sin\theta} \left(v_{\hat{\varphi}}^2 \frac{\partial f^{\rm eq}}{\partial v^{\hat{\theta}}} -  v^{\hat{\theta}} \frac{\partial (v^{\hat{\varphi}} f^{\rm eq})}{\partial v^{\hat{\varphi}}}\right) \Bigg].
 \label{eq:CE_df_sph}
\end{multline}

We are interested in the form of the viscous part $\sigma^{\hat{a}\hat{b}} = \tau^{\hat{a}\hat{b}} + p \delta^{\hat{a}\hat{b}}$ of the stress tensor, obtained from $\delta f$ via
\begin{equation}
 \sigma^{{\hat{a}}{\hat{b}}} = 
 m\int d\mathbf{v}\, \delta f \,v^{\hat{a}} v^{\hat{b}}.
\end{equation}
Multiplying Eq.~\eqref{eq:CE_df_sph} by $v^{\hat{a}} v^{\hat{b}}$ and  integrating with respect to the velocity space shows that the recovery of $\sigma_{{\hat{a}}{\hat{b}}}$ requires various moments of the equilibrium distribution, $f^{\rm eq}$, which we list below:
\begin{gather}
 \int d\mathbf{v}\, f^{\rm eq} = n, \quad 
 \int d\mathbf{v}\, f^{\rm eq} v^{\hat{a}} = n u^{\hat{a}}, \nonumber\\
 m\int d\mathbf{v}\, f^{\rm eq} v^{\hat{a}} v^{\hat{b}} = \rho u^{\hat{a}} u^{\hat{b}} + p \delta^{\hat{a}\hat{b}}, \nonumber\\
 m \int d\mathbf{v}\, f^{\rm eq} v^{\hat{a}} v^{\hat{b}} v^{\hat{c}} = \rho u^{\hat{a}} u^{\hat{b}} u^{\hat{c}} + p (u^{\hat{a}} \delta^{\hat{b}\hat{c}} + u^{\hat{b}} \delta^{\hat{c}\hat{a}} + u^{\hat{c}} \delta^{\hat{a}\hat{b}}).
\end{gather}
While the above relations are exactly satisfied by the $Q = 4$ quadrature, in the case when $Q = 3$, we find discrepancies for the third-order moment:
\begin{equation}
 m \sum_{\vb{k}} f_{\vb{k}}^{Q=3,{\rm eq}} v_{\vb{k}}^{\hat{a}} v_{\vb{k}}^{\hat{b}} v_{\vb{k}}^{\hat{c}} = p (u^{\hat{a}} \delta^{\hat{b}\hat{c}} + u^{\hat{b}} \delta^{\hat{c}\hat{a}} + u^{\hat{c}} \delta^{\hat{a}\hat{b}}),
\end{equation}
indicating that the $O(\vb{u}^3)$ term is not captured.  This omission is easily understood from the fact that the equilibrium distribution $f_{\vb{k}}^{Q=3,\mathrm{eq}}$, given in Eq.~\eqref{eq:feq_Q=3}, contains no terms cubic in the velocity $\vb{u}$. Nevertheless, as is well known within the LBM community, the missing contribution $O(\mathrm{Ma}^3)$ has a negligible impact in the low-Mach number regime.

In addition, the curvature term requires the recovery of the following moments of the velocity gradients of $f^{\rm eq}$:
\begin{align}
 \int d\mathbf{v} \frac{\partial f^{\rm eq}}{\partial v^{\hat{\theta}}} v_{\hat{\varphi}}^4 &=  \int d\mathbf{v} \frac{\partial (f^{\rm eq} v^{\hat{\varphi}})}{\partial v^{\hat{\varphi}}} v_{\hat{\theta}}^3 = 0, \nonumber\\
 m \int d\mathbf{v} \frac{\partial f^{\rm eq}}{\partial v^{\hat{\theta}}} v_{\hat{\varphi}}^3 v^{\hat{\theta}} &= -\rho u_{\hat{\varphi}}^3 - 3p u^{\hat{\varphi}}, \nonumber\\
 m \int d\mathbf{v} \frac{\partial f^{\rm eq}}{\partial v^{\hat{\theta}}} v_{\hat{\varphi}}^2  v_{\hat{\theta}}^2 &= m\int d\mathbf{v} \frac{\partial (f^{\rm eq} v^{\hat{\varphi}})}{\partial v^{\hat{\varphi}}} v^{\hat{\theta}} v_{\hat{\varphi}}^2 \nonumber\\
 &= -2\rho u_{\hat{\varphi}}^2 u^{\hat{\theta}} - 2p u^{\hat{\theta}}, 
 \nonumber\\
 m\int d\mathbf{v} \frac{\partial (f^{\rm eq} v^{\hat{\varphi}})}{\partial v^{\hat{\varphi}}} v_{\hat{\theta}}^2 v^{\hat{\varphi}} &= -\rho u^{\hat{\varphi}} u_{\hat{\theta}}^2 - p u^{\hat{\varphi}}.
\end{align}
Again, these relations are exactly satisfied in the case of the $Q = 4$ quadrature. When $Q = 3$, we find discrepancies in the following cases:
\begin{align}
 m \sum_{\vb{k}} \left(\frac{\partial f^{\rm eq}}{\partial v^{\hat{\theta}}}\right)_{\vb{k}}^{Q = 3;{\rm eq}} v_{k_\varphi}^3 v_{k_\theta}^2 &= -3p u^{\hat{\varphi}}, \\
 m \sum_{\vb{k}} \left(\frac{\partial f^{\rm eq}}{\partial v^{\hat{\theta}}}\right)_{\vb{k}}^{Q = 3;{\rm eq}} v_{k_\varphi}^2 v_{k_\theta}^2 &= -2p u^{\hat{\theta}}, \\
 m \sum_{\vb{k}} \left(\frac{\partial (f^{\rm eq} v^{\hat{\varphi}})}{\partial v^{\hat{\varphi}}}\right)_{\vb{k}}^{Q = 3;{\rm eq}} v_{k_\theta} v_{k_\varphi}^2 &= -2p u^{\hat{\theta}}, \\
 m \sum_{\vb{k}} \left(\frac{\partial (f^{\rm eq} v^{\hat{\varphi}})}{\partial v^{\hat{\varphi}}}\right)_{\vb{k}}^{Q = 3;{\rm eq}} v^2_{k_\theta} v_{k_\varphi} &= -p u^{\hat{\varphi}},
\end{align}
therefore again missing the $O(\vb{u}^3)$ terms. 

Summarizing the above discussion, we have $\sigma^{\hat{a}\hat{b}}_{Q = 4} = \sigma^{\hat{a}\hat{b}}_{\rm NS}$, with $\sigma^{\hat{a}\hat{b}}_{\rm NS}$ being the Navier-Stokes viscous stress tensor given in Eq.~\eqref{eq:viscous_stress} with $\zeta = \mu = \tau p$, while the viscous stress tensor obtained with $Q = 3$ deviates from this expression via
\begin{multline}
 \sigma^{\hat{a}\hat{b}}_{Q = 3} = \sigma^{\hat{a}\hat{b}}_{\rm NS} + \frac{\tau}{R \sin\theta}\left[\partial_\theta(\rho u^{\hat{a}} u^{\hat{b}} u^{\hat{\theta}}) + \partial_\varphi(\rho u^{\hat{a}} u^{\hat{b}} u^{\hat{\varphi}})\right] \\
 - \frac{\tau \cos\theta}{R \sin\theta} \left[\rho u_{\hat{\varphi}}^2(\delta^{\hat{a}}_{\hat{\theta}} u^{\hat{b}} + \delta^{\hat{b}}_{\hat{\theta}} u^{\hat{a}}) - \rho u^{\hat{\varphi}} u^{\hat{\theta}} (\delta^{\hat{a}}_{\hat{\varphi}} u^{\hat{b}} + \delta^{\hat{b}}_{\hat{\varphi}} u^{\hat{a}})\right].
\end{multline}
Due to the above inaccuracy, we consider the $Q = 4$ quadrature in the main text, however it is interesting to note that, aside from the shock wave problem considered in Sec.~\ref{sec:numerics:riemann}, the $Q = 3$ model should provide qualitatively satisfactory results for all tests considered in this paper.

\section{Solutions for Axis-symmetrical Flows}\label{app:axi}
In this appendix section, we provide full details on the analytic derivation of the solutions for the axisymmetric flows presented in Sec~\ref{sec:numerics}. The solution for the sound waves propagation, considered in Sec.~\ref{sec:numerics:sound}, is presented in Subsections~\ref{app:axi:sound}. The solution for the shear waves, considered in Sec.~\ref{sec:numerics:shear}, is presented in Subsec.~\ref{app:axi:shear}.

\subsection{Sound waves}\label{app:axi:sound}
We consider the propagation of longitudinal waves of infinitesimal amplitude along the polar ($\theta$) direction. We assume that the velocity along the azimuthal direction vanishes, $u^{\hat{\varphi}} = 0$. Writing 
\begin{equation}
 \rho = \rho_0 (1 + \delta \rho), \ 
 P = P_0(1 + \delta P),
\end{equation}
where $\rho_0$ and $P_0$ are the density and pressure of the background (undisturbed) fluid, and considering that $\delta \rho$, $\delta P$ and $u^{\hat{\theta}} \equiv \delta u^{\hat{\theta}}$ are infinitesimal perturbations, Eqs.~\eqref{eq:axi_continuity} and \eqref{eq:axi_NSE_uth} reduce to:
\begin{subequations}\label{eq:sound_eqs}
\begin{equation}
 \partial_t \delta \rho + \frac{\partial_\theta(\delta u^{\hat{\theta}} \sin\theta)}{R \sin \theta} = 0, \hfill 
 \label{eq:sound_continuity}
\end{equation}
\begin{equation} 
  \partial_t \delta u^{\hat{\theta}} + \frac{c_s^2}{R}  \partial_\theta \delta \rho 
  = \frac{\nu + \nu_v}{R^2} \frac{\partial}{\partial \theta} \left[\frac{\partial_\theta(\sin\theta \delta u^{\hat{\theta}})}{\sin\theta}\right] + \frac{2\nu}{R^2} \delta u^{\hat{\theta}},
  \label{eq:sound_uth}
\end{equation}
\end{subequations}
where $\nu = \eta / \rho$ and $\nu_v = \zeta / \rho$ are the kinematic shear and bulk viscosities and we took into account that the pressure perturbations can be linked to the density perturbations via
\begin{equation}
 \delta P= \frac{\rho_0}{P_0} \frac{\partial P}{\partial \rho} \delta \rho,
\end{equation}
with $\partial P / \partial \rho = c_s^2$ being the speed of sound squared.

{\bf Inviscid fluid.} We first consider the case of a perfect fluid, corresponding to vanishing kinematic viscosities $\nu = \nu_v = 0$. In this case, Eq.~\eqref{eq:sound_uth} reduces to
\begin{equation}
 \partial_t \delta u^{\hat{\theta}} + \frac{c_s^2}{R}  \partial_\theta \delta \rho = 0. \label{eq:sound_uth_inv}   
\end{equation}
Combining Eqs.~\eqref{eq:sound_continuity} and \eqref{eq:sound_uth_inv}, we arrive at the single differential equation 
\begin{equation}\label{eq:sound_wave_on_sphere_eq}
    \frac{\partial^2 \delta u^{\hat{\theta}}}{\partial t^2} - \frac{c^2_{s}}{R^2} \frac{\partial}{\partial \theta}\left[\frac{1}{\sin{\theta}} 
    \frac{\partial}{\partial \theta} \left(\delta u^{\hat{\theta}} \sin\theta \right) \right] = 0.
\end{equation}

The above equation can be solved using the separation of variables of the form 
\begin{equation}
 \delta u^{\hat{\theta}} \to \frac{U_n(t) F_n(\theta)}{\sin\theta}, 
\end{equation}
where $n$ is used to identify the different possible solutions of Eq.~\eqref{eq:sound_wave_on_sphere_eq}. This leads to a system of two decoupled equations in $t$ and $\theta$:
\begin{subequations}
\begin{align}
    \dv[2]{U_n}{t} + \lambda^2_n \dfrac{c^2_s}{R^2}U_n &= 0,  \label{eq:sound_dUn}\\
    \sin{\theta} \dv{\theta}\left(\dfrac{1}{\sin{\theta}} \pdv{F_n}{\theta}\right) + \lambda^2_n F_n &= 0.\label{eq:sound_dFndth}
\end{align}
\end{subequations}
We easily recognize that \eqref{eq:sound_dUn} has the form of a harmonic oscillator equation with the solution:
\begin{equation}\label{eq:sound_Un_sol}
 U_n(t) = u_n \cos(\omega_n t + \varphi_n),
\end{equation}
where $u_n$ and $\varphi_n$ represent arbitrary integration constants, while the angular frequency obeys
\begin{equation}
 \omega_n = \frac{\lambda_n c_s}{R}. \label{eq:sound_omegan}
\end{equation}

For the angular equation, we seek even and odd solutions with respect to $\theta \rightarrow \pi - \theta$,
\begin{equation}
 F^e_n(\theta) \equiv f^e_n(\zeta), \quad 
 F^o_n(\theta) \equiv \cos\theta\, f^o_n(\zeta),
\end{equation}
where we take $\zeta = \cos 2\theta$ to be the argument of the reduced functions $f^{e/o}_n$. In this case, Eq.~\eqref{eq:sound_dFndth} becomes 
\begin{align}
 (1 - \zeta^2) \frac{\partial^2 f^e_n}{\partial \zeta^2} + \frac{1}{2}(1 - \zeta) \frac{\partial f^e_n}{\partial \zeta} + \frac{\lambda_{e;n}^2}{4} f^e_n &= 0,
 \label{eq:sound_dfn}\\
 (1 - \zeta^2) \frac{\partial^2 f^o_n}{\partial \zeta^2} + \frac{3}{2}(1 - \zeta) \frac{\partial f^o_n}{\partial \zeta} + \frac{\lambda_{o;n}^2}{4} g_n &= 0.
 \label{eq:sound_dgn}
\end{align}
The regular solutions of the above equations can be written in terms of the Jacobi polynomials $P_n^{(\alpha,\beta)}(x)$, which satisfy
\begin{multline}\label{eq:Jacobi}
 \bigg[(1-x^2) \frac{d^2}{dx^2} + [\beta - \alpha - (\alpha + \beta + 2) x] \frac{d}{dx} \\ + n(n + \alpha + \beta + 1)\bigg] P_n^{(\alpha,\beta)}(x) = 0.
\end{multline}
For future reference, we give below the series expansion for the Jacobi polynomials:
\begin{multline}
 P^{(\alpha,\beta)}_n(\zeta) = \frac{\Gamma(\alpha + n + 1)}{n! \Gamma(\alpha + \beta + n + 1)} \\
 \times \sum_{m = 0}^n \binom{n}{m} \frac{\Gamma(\alpha + \beta + n + m + 1)}{\Gamma(\alpha + m + 1)} \left(\frac{\zeta - 1}{2}\right)^m,
 \label{eq:Jacobi_series}
\end{multline}
as well as the orthogonality relation 
\begin{multline}
 \int_{-1}^1 dx\, (1-x)^\alpha (1+x)^\beta P_m^{(\alpha,\beta)}(x) P_n^{(\alpha,\beta)}(x) \\
 = \frac{2^{\alpha+\beta+1}}{2n+\alpha+\beta+1} 
 \frac{\Gamma(n + \alpha + 1) \Gamma(n + \beta + 1)}{\Gamma(n + \alpha + \beta + 1)n!} \delta_{nm}.
 \label{eq:Jacobi_ortho}
\end{multline}

Comparing Eqs.~\eqref{eq:sound_dfn}, \eqref{eq:sound_dgn} and \eqref{eq:Jacobi}, it can be seen that the angular functions $F^{e/o}_n$ are given by
\begin{subequations}\label{eq:sound_Fn_aux}
\begin{align}
 F^e_n(\theta) &= \mathcal{N}_{e;n} P^{(-1, -1/2)}_n(\cos2\theta), \\
 F^o_n(\theta) &= \mathcal{N}_{o;n} \cos\theta P^{(-1,1/2)}_n(\cos2\theta),
\end{align}
\end{subequations}
where $n = 0, 1, 2, \dots$ must be a non-negative integer. The eigenvalues $\lambda^{e/o}_n$ are related to $n$ via
\begin{equation}
 \lambda^e_n = \sqrt{2n(2n - 1)}, \quad 
 \lambda^o_n = \sqrt{2n(2n + 1)}.
 \label{eq:sound_lambda_aux}
\end{equation}
The normalization constants $\mathcal{N}^{e/o}_n$ appearing in Eqs.~\eqref{eq:sound_Fn_aux} must be found by imposing unit norm with respect to a suitable inner product for the acoustic modes (``ac''),
\begin{equation}
 \langle F^e_n, F^e_{n'} \rangle_{\rm ac} = \langle F^o_n, F^o_{n'} \rangle_{\rm ac} = \delta_{nn'},
\end{equation}
with $\langle F^e_n, F^o_{n'} \rangle_{\rm ac} = 0$ automatically, by symmetry considerations. The relevant inner product can be found as follows. 

First, we multiply Eq.~\eqref{eq:sound_dFndth} by $F_{n'} / \sin\theta$, with $F_{n'}$ being also a solution of Eq.~\eqref{eq:sound_dFndth}. From the result, we subtract an equivalent expression with $n$ and $n'$ interchanged, leading to
\begin{equation}
 \frac{\partial}{\partial \theta} \left[\frac{(F_{n'} \overleftrightarrow{\partial_\theta} F_n)}{\sin\theta}  \right] + \frac{\lambda_n^2 - \lambda_{n'}^2}{\sin\theta} F_{n'}(\theta) F_n(\theta) = 0,
\end{equation}
where $f \overleftrightarrow{\partial_\theta} g = f\partial_\theta g - (\partial_\theta f) g$ is the billateral derivative.
Integrating the above with respect to $\theta$ between $0$ and $\pi$ and assuming that the term inside the square brackets vanishes sufficiently fast around the spherical poles at $\theta = 0$ and $\pi$, we arrive at
\begin{equation}
 (\lambda_n^2 - \lambda_{n'}^2) \langle F_n, F_{n'} \rangle_{\rm ac} = 0, 
\end{equation}
where 
\begin{equation}
 \langle \varphi, \chi \rangle_{\rm ac} = \int_{0}^\pi \frac{d\theta}{\sin\theta} \varphi(\theta) \chi(\theta).
\end{equation}

Substituting $\alpha = -1$ and $\beta = -1/2$ into Eq.~\eqref{eq:Jacobi_ortho} and taking into account that the interval $[-1,1]$ for $x = \cos 2\theta$ covers only half of the sphere, we have 
\begin{multline}
\int_0^\pi \frac{d\theta}{\sin\theta} P_m^{(-1,-1/2)}(\cos 2\theta) P_n^{(-1,-1/2)}(\cos 2\theta) \\ = \frac{2n - 1}{n(4n-1)} \delta_{nm}, \nonumber
\end{multline}
\begin{multline}
 \int_0^\pi \frac{d\theta}{\sin\theta} \cos^2\theta\, P_m^{(-1,1/2)}(\cos 2\theta) P_n^{(-1,1/2)}(\cos 2\theta) \\
 = \frac{2n + 1}{n(4n+1)} \delta_{nm},\nonumber
\end{multline}
such that the normalization constants are 
\begin{equation}
 \mathcal{N}^e_n = \sqrt{\frac{n(4n-1)}{2n-1}}, \quad 
 \mathcal{N}^o_n = \sqrt{\frac{n(4n+1)}{2n+1}}.
\end{equation}
From the above, it can be seen that the solutions with $n = 0$ remain unviable since they cannot be normalized. Taking into account that $P^{(\alpha,\beta)}_0(z) = 1$, the velocity profile corresponding to the $n = 0$ solutions is either $u^{\hat{\theta}} \sim 1 / \sin\theta$ (even) or $u^{\hat{\theta}} \sim \cos\theta / \sin\theta$ (odd), both versions diverging near the poles at $\theta = 0$ and $\pi$.

Thus, the even and odd solutions $F^{e/o}_n(\theta)$ of the angular problem for the propagation of acoustic (sound) modes read:
\begin{subequations}\label{eq:sound_Fn}
\begin{align}
 F^e_n(\theta) &= (-1)^n \sqrt{\frac{n(4n-1)}{2n-1}} P^{(-1,-1/2)}_n(\cos 2\theta), \label{eq:sound_Fen} \\ 
 F^o_n(\theta) &= (-1)^n \cos\theta \sqrt{\frac{n(4n+1)}{2n+1}} P^{(-1,1/2)}_n(\cos 2\theta),  \label{eq:sound_Fon}
\end{align}
\end{subequations}
with the corresponding eigenvalues given in Eq.~\eqref{eq:sound_lambda_aux}. Before listing the full solution $u^{\hat{\theta}}(t,\theta)$, we discuss below the inclusion of dissipative effects.

{\bf Dissipative fluid.} In order to incorporate the effect of the dissipative term on the right-hand side of Eq.~\eqref{eq:sound_uth}, we first write out explicitly the time dependence of our normal modes:
\begin{equation}
 \delta \rho = \sum_n e^{-i \alpha_n t} \widetilde{\delta \rho}_n(\theta), \quad 
 u^{\hat{\theta}} = \sum_n e^{-i \alpha_n t} \tilde{u}^{\hat{\theta}}_n(\theta),
\end{equation}
where $\widetilde{\delta \rho}_n$ and $\tilde{u}^{\hat{\theta}}_n$ depend only on the polar coordinate $\theta$. The continuity equation \eqref{eq:sound_continuity} gives 
\begin{subequations}\label{eq:sound_diss_eqs}
\begin{equation}
 \widetilde{\delta \rho}_n = -\frac{i}{R \alpha_n \sin\theta} \frac{\partial(\tilde{u}^{\hat{\theta}}_n \sin\theta)}{\partial \theta}. 
\end{equation}
Plugging this into Eq.~\eqref{eq:sound_uth} gives 
\begin{multline}
 \left(-i R^2 \alpha_n + 2\nu\right) \tilde{u}^{\hat{\theta}}_n = \\
 \left(\frac{i c_s^2}{\alpha_n} + \nu + \nu_v\right) 
 \frac{\partial}{\partial \theta} \left[\frac{1}{\sin\theta} \frac{\partial(\tilde{u}^{\hat{\theta}}_n\sin\theta)}{\partial \theta} \right].
 \label{eq:sound_uth_diss}
\end{multline}
\end{subequations}
The above equation looks just like Eq.~\eqref{eq:sound_dFndth} appearing in the case of the perfect fluid, which admits the mode solutions $\tilde{u}^{\hat{\theta}} \to F_n(\theta) / \sin\theta$. 

Considering $\tilde{u}^{\hat{\theta}} \to F_n(\theta) / \sin\theta$ in Eq.~\eqref{eq:sound_uth_diss} and using Eq.~\eqref{eq:sound_dFndth} to eliminate the differential operator leads to the following equation for the frequencies $\alpha_n$:
\begin{equation}
 \alpha^2_n + \frac{i \nu \lambda_n^2}{R^2} \left(\frac{\lambda_n^2 - 2}{\lambda_n^2} + \frac{\nu_v}{\nu}\right) - \frac{c_s^2 \lambda_n^2}{R^2} = 0.
\end{equation}
In general, we have two solutions given by
\begin{gather}
 \alpha_n^\pm = 
 -i \gamma_n \pm \omega_n(\nu), \quad 
 \gamma_n = \frac{\nu \lambda_n^2}{2R^2} \left(\frac{\lambda_n^2 - 2}{\lambda_n^2} + \frac{\nu_v}{\nu}\right), \nonumber\\
 \omega_n(\nu) = \sqrt{\omega^2_n(0) - \gamma_n^2}, 
 \label{eq:sound_damping_coeffs}
\end{gather}
where $\omega_n(0) = c_s \lambda_n / R$ is the frequency encountered in the dissipationless case [see Eq.~\eqref{eq:sound_omegan}]. In the absence of dissipation ($\nu = \nu_v = 0$), we recover the frequency encountered in the inviscid case, $\alpha_n \to \omega_n(0) = \pm c_s \lambda_n / R$. It is remarkable that in the absence of bulk viscosity ($\nu_v = 0$) the first even mode for which $\lambda^e_1 = \sqrt{2}$ is dissipationless. 

The general solution for the sound wave propagation problem reads
\begin{equation}
 u^{\hat{\theta}}(t,\theta) = \frac{1}{\sin\theta} \sum_{n = 1}^\infty \left[U^e_n(t) F^e_n(\theta) + 
 U^o_n(t) F^o_n(\theta) \right],
 \label{eq:sound_sol}
\end{equation}
where the time-dependent amplitudes $U^*_n(t)$ ($* \in \{e,o\}$) are given by
\begin{equation}
 U^*_n(t) = u^*_n e^{-\gamma^*_n t} \cos[\omega^*_n(\nu) t + \varphi^*_n],
 \label{eq:sound_ampl}
\end{equation}
where the dissipation coefficient $\gamma^*_n$ and the angular velocity $\omega^*_n$ are given in Eqs.~\eqref{eq:sound_damping_coeffs}.
The constants $u^*_n$ and $\varphi^*_n$ can be obtained from the initial density gradient, $\delta \rho'_0 \equiv \partial_\theta \delta \rho(t=0,\theta)$, and initial velocity profile, $u^{\hat{\theta}}_0 \equiv u^{\hat{\theta}}(t = 0, \theta)$, via 
\begin{subequations}\label{eq:sound_un}
\begin{align}
 u^*_n (\gamma_n^* \cos\varphi_n^* - \omega_n^* \sin\varphi_n^*) &= 
 -\frac{c_s^2}{R} \int_0^\pi d\theta\, \delta \rho'_0 F^*_n(\theta),\label{eq:sound_unsin}  \\
 u^*_n \cos\varphi_n^* &= 
 \int_0^\pi d\theta\, u^{\hat{\theta}}_0 F^*_n(\theta).\label{eq:sound_uncos}
\end{align}
\end{subequations}

\subsection{Shear waves}\label{app:axi:shear}

We now consider the dynamics of a shear wave $u^{\hat{\varphi}} \equiv u^{\hat{\varphi}}(t,\theta)$, governed by the $\varphi$ component of the Cauchy equation \eqref{eq:axi_NSE_uph}. Considering $u^{\hat{\varphi}} \to \delta u^{\hat{\varphi}}$ as an infinitesimal perturbation gives
\begin{equation}
 \frac{\partial}{\partial t} \left(\frac{u^{\hat{\varphi}}}{\sin\theta}\right) = \frac{\nu}{R^2 \sin^3\theta} \frac{\partial}{\partial \theta} \left[\sin^3 \theta \frac{\partial}{\partial \theta} \left(\frac{u^{\hat{\varphi}}}{\sin\theta}\right)\right].
 \label{eq:shear_duphidt}
\end{equation}
To solve this equation, we use separation of variables and consider normal modes of the form
\begin{equation}
 u^{\hat{\varphi}} \rightarrow \sin\theta e^{-\kappa_n t} G_n(\theta),
\end{equation}
such that $G_n(\theta)$ satisfies
\begin{equation}
 \frac{1}{\sin^3\theta} \frac{\partial}{\partial \theta} \left(\sin^3 \theta \frac{\partial G_n}{\partial \theta} \right) + \chi_n^2 G_n = 0, \label{eq:shear_dGn}
\end{equation}
where $\chi_n = R \sqrt{\kappa_n / \nu}$ represent the eigenvalues of the differential operator appearing above.

We seek for odd and even solutions of Eq.~\eqref{eq:shear_dGn}  with respect to $ \theta \rightarrow \pi - \theta$, $G^e_n$ and $G^o_n$, defined as
\begin{equation}
 G^e_n(\theta) \equiv g^e_n(\zeta), \quad 
 G^o_n(\theta) \equiv \cos\theta g^o_n(\zeta),
 \label{eq:shear_FG}
\end{equation}
where we introduced the variable $\zeta = \cos(2\theta)$. The functions $g^{e/o}_n$ satisfy the following differential equations:
\begin{align}
 & (1 - \zeta^2) \frac{\partial^2 g^e_n}{\partial \zeta^2} -\frac{1}{2} (3 + 5 \zeta) \
 \frac{\partial g^e_n}{\partial \zeta} + \frac{\chi^2_{e;n}}{4} g^e_n = 0,\label{eq:shear_gen} \\
 & (1 - \zeta^2) \frac{\partial^2 g^o_n}{\partial \zeta^2} - \frac{1}{2} (1 + 7 \zeta) \
 \frac{\partial g^o_n}{\partial \zeta} + \left( \frac{\chi^2_{o;n}}{4} - 1 \right) g^o_n = 0.\label{eq:shear_gon}
\end{align}
Comparing the above to the differential equation \eqref{eq:Jacobi} satisfied by the Jacobi polynomials $P^{(\alpha,\beta)}_n(x)$, 
we see that $(\alpha,\beta) = (1,-1/2)$ and $(1,1/2)$ for the even and odd solutions, respectively, such that 
\begin{subequations}\label{eq:shear_geo_aux}
\begin{align}
 g^e_n &= \mathcal{C}^e_n P^{(1,-1/2)}_n(\cos 2\theta), &
 \chi_{e;n} &= \sqrt{2n(2n + 3)}, \label{eq:shear_ge_aux}\\
 g^o_n &= \mathcal{C}^o_n P^{(1,-1/2)}_n(\cos 2\theta), &
 \chi_{o;n} &= \sqrt{2(2n+1)(n+2)}, 
 \label{eq:shear_go_aux}
\end{align}
\end{subequations}
with $n = 0, 1, 2, \dots$ being a non-negative integer and $\mathcal{C}^{e/o}_n$ normalization constants.

In order to fix the normalization constants $\mathcal{C}^{e/o}_n$ appearing in Eqs.~\eqref{eq:shear_geo_aux}, we will employ a suitable inner product $\langle \varphi, \chi \rangle_{\rm sh}$, such that
\begin{equation}
 \langle G^e_n, G^e_{n'} \rangle_{\rm sh} = \langle G^o_n, G^o_{n'} \rangle_{\rm sh} = \delta_{n,n'}, 
\end{equation}
while $\langle G^e_n, G^o_{n'} \rangle_{\rm sh} = 0$ by symmetry considerations.
The inner product compatible with the differential equation \eqref{eq:shear_dGn} for a generic solution $G_n$ can be obtained by multiplying this equation by $G_{n'} \sin^3\theta$, with $G_{n'}$ being another solution of the same equation. Subtracting an equivalent relation with $n \leftrightarrow n'$ gives 
\begin{equation}
 \partial_\theta \left[\sin^3 \theta \left(G_{n'} \overleftrightarrow{\partial_\theta} G_n \right) \right] + (\chi_n^2 - \chi_{n'}^2) \sin^3\theta G_{n'} G_n = 0.
\end{equation}
Integrating the above equation with respect to $\theta$ between $0$ and $\pi$ provides the natural definition for the inner product, i.e.
\begin{gather}
 (\chi_n^2 - \chi_{n'}^2) \langle G_n, G_{n'} \rangle_{\rm sh} = 0, \nonumber\\
 \langle \varphi, \chi \rangle_{\rm sh} \equiv \int_0^\pi d\theta \, \sin^3 \theta \varphi(\theta) \chi(\theta).
\end{gather}
On the other hand, the orthogonality relation \eqref{eq:Jacobi_ortho} for the Jacobi polynomials for the cases $(\alpha,\beta) = (1,-1/2)$ and $(1,1/2)$ reads as follows:
\begin{multline}
 \int_0^\pi d\theta \,\sin^3 \theta\, 
 P^{(1,-\frac{1}{2})}_n (\cos2\theta) 
 P^{(1,-\frac{1}{2})}_m(\cos2\theta) \\ 
 =\frac{4(n+1)\delta_{nm}}{(2n+1)(4n+3)},
\end{multline}
\begin{multline}
 \int_0^\pi d\theta \,\sin^3 \theta\, \cos^2\theta 
 P^{(1,\frac{1}{2})}_n (\cos2\theta) 
 P^{(1,\frac{1}{2})}_m(\cos2\theta) \\ 
 =\frac{4(n+1)\delta_{nm}}{(2n+3)(4n+5)}.
\end{multline}
Thus, the normalization constants $\mathcal{C}^{e/o}_n$ satisfy
\begin{subequations}
\begin{align}
 \mathcal{C}^e_n = \sqrt{\frac{2(n+\frac{1}{2})(n+\frac{3}{4})}{n+1}}, \quad
 \mathcal{C}^o_n = \sqrt{\frac{2(n+\frac{3}{2})(n+\frac{5}{4})}{n+1}}.
\end{align}
\end{subequations}

In summary, the mode solutions of Eq.~\eqref{eq:shear_dGn} for the angular part of the shear waves problem read 
\begin{subequations}\label{eq:shear_Gn}
\begin{align}
 G^e_n(\theta) &= \sqrt{\frac{(2n+1)(4n+3)}{4(n+1)}} P^{(1,-\frac{1}{2})}_n(\cos2\theta), \label{eq:shear_Gen} \\ 
 G^o_n(\theta) &= \sqrt{\frac{(2n+3)(4n+5)}{4(n+1)}} \cos\theta \, P^{(1,\frac{1}{2})}_n(\cos2\theta),\label{eq:shear_Gon}
\end{align}
\end{subequations}
while their corresponding eigenvalues are
\begin{equation}
 \chi^e_n = \sqrt{4n(n + \tfrac{3}{2})}, \quad
 \chi^o_n = \sqrt{2(2n+1)(n+2)}.
 \label{eq:shear_chin}
\end{equation}
The general solution for the shear wave problem can be written as
\begin{equation}
 u^{\hat{\varphi}}(t,\theta) = \sin\theta \sum_{n = 0}^\infty \left[V_n^e(t) G^e_n(\theta) + V_n^o(t) G^o_n(\theta)\right],
 \label{eq:shear_uphi_sol}
\end{equation}
where the time-dependent amplitudes are
\begin{equation}
 V_n^e(t) = v^e_n e^{-\kappa^e_n t}, \quad 
 V_n^o(t) = v^o_n e^{-\kappa^o_n t}.
 \label{eq:shear_Vn}
\end{equation}
The damping coefficients $\kappa^{e/o}_n$ read:
\begin{subequations}\label{eq:shear_kappan}
\begin{align}
 \kappa^e_n &= \frac{\nu}{R^2}  \chi_{e;n}^2 = \frac{\nu}{R^2} 2n (2n + 3), \label{eq:shear_kappaen}\\
 \kappa^o_n &= \frac{\nu}{R^2}  \chi_{o;n}^2 = \frac{2\nu}{R^2}  (2n+1)(n+2).
 \label{eq:shear_kappaon}
\end{align}
\end{subequations}
The constant amplitudes $v^{e/o}_n$ appearing in Eqs.~\eqref{eq:shear_Vn} can be obtained from the initial velocity profile $u^{\hat{\varphi}}_0(\theta) \equiv u^{\hat{\varphi}}(t = 0, \theta)$ via
\begin{equation}
 v^{e/o}_n = \int_0^\pi d\theta\, \sin^2\theta \, u^{\hat{\varphi}}_0(\theta) G_n^{e/o}(\theta).
 \label{eq:shear_veon}
 \end{equation}

\section{Isotropy test}\label{app:rot}
In this appendix section, we test the isotropy preservation of our proposed vielbein lattice Boltzmann scheme by considering the axisymmetric flows described in Sec. \ref{app:axi} in a rotated reference frame. To be specific, we consider a new coordinate frame given by the spherical coordinates ($\theta', \varphi'$), obtained by applying a clockwise rotation of angle $\alpha$ with respect to the $y$ axis on the initial coordinate system. With respect to the initial coordinate axes $Ox$, $Oy$ and $Oz$, having unit vectors $\mathbf{i}$, $\mathbf{j}$ and $\mathbf{k}$, respectively, the new axes have the unit vectors
\begin{equation}
 \mathbf{i}' = \mathbf{i} \cos\alpha + \mathbf{k} \sin\alpha, \quad 
 \mathbf{j}' = \mathbf{j}, \quad 
 \mathbf{k}' = -\mathbf{i} \sin\alpha + \mathbf{k} \cos\alpha.
\end{equation}
Consider a point $P$ on the sphere, having the position vector $\mathbf{x}(P) = x \mathbf{i} + y \mathbf{j} + z \mathbf{k}$ expressed with respect to the original coordinate system. In the new coordinate system, the same vector can be decomposed as $\mathbf{x}(P) = x' \mathbf{i}' + y'\mathbf{j}' + z' \mathbf{k}'$, with 
\begin{equation}
 \begin{pmatrix}
  x' \\ y' \\ z'
 \end{pmatrix} = R_2(\alpha) \begin{pmatrix}
  x \\ y \\ z 
 \end{pmatrix}, \ 
 R_2(\alpha) = \begin{pmatrix}
  \cos\alpha & 0 & \sin\alpha \\
  0 & 1 & 0 \\
  -\sin\alpha & 0 & \cos\alpha
 \end{pmatrix}.
\end{equation}
The above relations allow the spherical coordinates $(\theta, \varphi)$ and $(\theta', \varphi')$ with respect to the original and rotated coordinate frames to be related via
\begin{subequations}\label{eq:transf}
\begin{align}
 \sin \theta' \cos\varphi' &= \cos\alpha \sin\theta \cos\varphi + \sin\alpha \cos\theta,\\
 \sin \theta' \sin\varphi' &= \sin \theta \sin\varphi, \\
 \cos\theta' &= -\sin\alpha \sin\theta \cos\varphi + \cos\alpha \cos\theta.
\end{align}    
\end{subequations}

The inverse transformation reads:
\begin{subequations} \label{eq:transf_inv}  
\begin{align} 
 \sin \theta \cos\varphi &= \cos\alpha \sin\theta' \cos\varphi' - \sin\alpha \cos\theta,\\
 \sin \theta \sin\varphi &= \sin \theta' \sin\varphi', \\
 \cos\theta &= \sin\alpha \sin\theta' \cos\varphi' + \cos\alpha \cos\theta'.
\end{align}
\end{subequations}

The above relations are sufficient to relate a scalar field, e.g., the density, between the two frames:
\begin{equation}
 \rho'(\theta', \varphi') = \rho(\theta, \varphi).
\end{equation}
When considering the components of a vector field, e.g., the velocity, we employ the relation
\begin{equation}
 \begin{pmatrix}
  u^{\theta'} \\ u^{\varphi'}
 \end{pmatrix} = 
 \begin{pmatrix}
  \partial \theta' / \partial \theta & 
  \partial \theta' / \partial \varphi \\
  \partial \varphi' / \partial \theta & \partial \varphi' / \partial \varphi
 \end{pmatrix} \begin{pmatrix}
  u^\theta \\ u^\varphi
 \end{pmatrix}.
\end{equation}
Using the relations in Eqs.~\eqref{eq:transf}, we arrive at 
\begin{widetext}
\begin{align}
 \begin{pmatrix}
  u'^{\hat{\theta}'} \\ u'^{\hat{\varphi'}}
 \end{pmatrix} &= 
 \frac{1}{\sin\theta} \begin{pmatrix}
  \cos\alpha \sin\theta' - \sin\alpha \cos\theta'\cos\varphi' & -\sin\alpha \sin\varphi' \\ 
  \sin\alpha \sin\varphi' & \cos\alpha \sin\theta' - \sin\alpha \cos\theta'\cos\varphi'
 \end{pmatrix} \begin{pmatrix}
  u^{\hat{\theta}} \\ u^{\hat{\varphi}}
 \end{pmatrix} \nonumber\\
 &= 
 \frac{1}{\sin\theta'} \begin{pmatrix}
  \cos\alpha \sin\theta + \sin\alpha \cos\theta \cos\varphi & -\sin\alpha \sin\varphi \\ 
  \sin\alpha \sin\varphi & \cos\alpha\sin\theta + \sin\alpha \cos\theta \cos\varphi
 \end{pmatrix} \begin{pmatrix}
  u^{\hat{\theta}} \\ u^{\hat{\varphi}}
 \end{pmatrix},
 \label{eq:transf_u}
\end{align}
where we used the relations $u^{\hat{\theta}} = u^{\theta}$ and $u^{\hat{\varphi}} = \sin\theta \, u^\varphi$ between the tetrad and coordinate components of the velocity vector field, while $\sin\theta$ and $\sin\theta'$ can be obtained via 
\begin{align}
 \sin^2\theta &= \sin^2\theta' \sin^2 \varphi' + (\cos\alpha \sin\theta' \cos\varphi' - \sin\alpha \cos\theta')^2, \nonumber\\
 \sin^2\theta' &= \sin^2\theta \sin^2 \varphi + (\cos\alpha \sin\theta \cos\varphi + \sin\alpha \cos\theta)^2.
\end{align}
The inverse relations read
\begin{align}
 \begin{pmatrix}
  u^{\hat{\theta}} \\ u^{\hat{\varphi}}
 \end{pmatrix} &= 
 \frac{1}{\sin\theta'} \begin{pmatrix}
  \cos\alpha \sin\theta + \sin\alpha \cos\theta \cos\varphi & \sin\alpha \sin\varphi \\ 
  -\sin\alpha \sin\varphi' & \cos\alpha \sin\theta + \sin\alpha \cos\theta \cos\varphi
 \end{pmatrix} \begin{pmatrix}
  u'^{\hat{\theta'}} \\ u'^{\hat{\varphi'}}
 \end{pmatrix} \nonumber\\
 &= 
 \frac{1}{\sin\theta} \begin{pmatrix}
  \cos\alpha \sin\theta' - \sin\alpha \cos\theta' \cos\varphi' & \sin\alpha \sin\varphi' \\ 
  -\sin\alpha \sin\varphi' & \cos\alpha\sin\theta' - \sin\alpha \cos\theta' \cos\varphi'
 \end{pmatrix} \begin{pmatrix}
  u'^{\hat{\theta}'} \\ u'^{\hat{\varphi}'}
 \end{pmatrix}.
 \label{eq:transf_u_inv}
\end{align}

Let us now consider the problem of extracting the amplitudes $U^{e/o}_n$ and $V^{e/o}_n$ of the harmonics in the sound and shear wave setups. These amplitudes are extracted in the frame where the flow is axisymmetric, as shown in Eqs.~\eqref{eq:sound_ampl} and \eqref{eq:shear_Vn}, via the formulas
\begin{align}
 U^{e/o}_n(t) &= \int_0^{2\pi} \frac{d\varphi}{2\pi} \int_0^\pi d\theta\, u^{\hat{\theta}}(t, \theta) F^{e/o}_n(\theta), \\
 V^{e/o}_n(t) &= \int_0^{2\pi} \frac{d\varphi}{2\pi} \int_0^\pi d\theta\,\sin^2\theta\, u^{\hat{\varphi}}(t, \theta) G^{e/o}_n(\theta),
 \label{eq:rot_UV_aux}
\end{align}
where we inserted an integration with respect to $\varphi$. We now switch the integration variables $d\Omega = \sin\theta d\theta d\varphi$ to $d\Omega' = \sin\theta' d\theta' d\varphi'$, where the Jacobian is unity as $d\Omega$ is an invariant integration measure (this can be checked by an explicit calculation). Thus, we arrive at the expressions
\begin{align}
 U^{e/o}_n &= \int_0^{2\pi} \frac{d\varphi'}{2\pi} \int_0^\pi \frac{d\theta' \sin\theta'}{\sin^2\theta} 
 [(\cos\alpha \sin\theta' - \sin\alpha \cos\theta' \cos\varphi') u'^{\hat{\theta}'} + \sin\alpha \sin\varphi' u'^{\hat{\varphi}'}] F^{e/o}_n(\theta), \label{eq:rot_U}\\
 V^{e/o}_n &= \int_0^{2\pi} \frac{d\varphi'}{2\pi} \int_0^\pi d\theta' \sin\theta' 
 [-\sin\alpha \sin\varphi' u'^{\theta'} + (\cos\alpha \sin\theta' - \sin\alpha \cos\theta' \cos\varphi') u'^{\hat{\varphi}'}] G^{e/o}_n(\theta),
 \label{eq:rot_V}
\end{align}
\end{widetext}
where $\theta$ and $\varphi$ can be obtained from Eq.~\eqref{eq:transf_inv}. 
\begin{figure}[tb]
    \includegraphics[width=0.99\columnwidth]{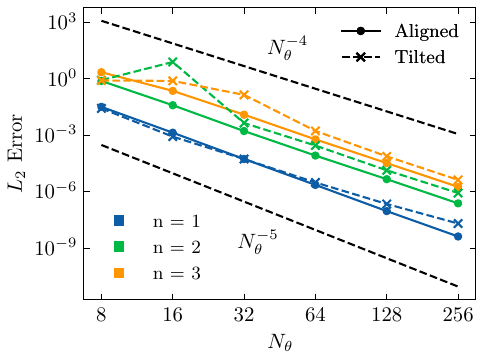}
    \caption{Convergence test for the sound wave solution. The L2 error evaluated between the analytical 
             and numerical solution for the $n=1,2,3$ modes using the \ac{weno5} scheme is plotted against 
             the grid size $N_\theta$. Continuous lines refer to data from simulations where the sound wave is aligned
             to the grid, dotted lines are for data where the initial condition is rotated clockwise by an angle $\alpha = - \pi / 2$. 
             In the axisymmetric case, we took $N_\varphi = 1$, while in the case of the tilted grid, we employed $N_\varphi = N_\theta$.
    }
    \label{fig:isotropy_test}
\end{figure}

We now specialize the above results for the case when $\alpha = \pi / 2$, which was considered in Sec.\ref{sec:numerics:riemann}. The transformation equations \eqref{eq:transf} and \eqref{eq:transf_inv} reduce to:
\begin{equation}
 \begin{pmatrix}
 \sin \theta' \cos\varphi' \\
 \sin \theta' \sin\varphi' \\
 \cos\theta' 
 \end{pmatrix} = \begin{pmatrix}
  \cos \theta \\ \sin\theta \sin\varphi \\ -\sin\theta\cos\varphi
 \end{pmatrix},
 \label{eq:transf_90}
\end{equation}
while the velocity components are related via
\begin{align}
 \begin{pmatrix}
  u'^{\hat{\theta}'} \\ u'^{\hat{\varphi'}}
 \end{pmatrix} &= 
 -\frac{\cos\varphi'}{\sin\theta} \begin{pmatrix}
  \cos\theta' & \tan\varphi' \\ 
  -\tan\varphi' & \cos\theta' 
 \end{pmatrix} \begin{pmatrix}
  u^{\hat{\theta}} \\ u^{\hat{\varphi}}
 \end{pmatrix}, \label{eq:transf_u_90}\\
 \begin{pmatrix}
  u^{\hat{\theta}} \\ u^{\hat{\varphi}}
 \end{pmatrix} &= 
 \frac{\cos\varphi}{\sin\theta'} \begin{pmatrix}
  \cos\theta & \tan\varphi \\ 
  -\tan\varphi & \cos\theta 
 \end{pmatrix} \begin{pmatrix}
  u'^{\hat{\theta}'} \\ u'^{\hat{\varphi}'}
 \end{pmatrix},
 \label{eq:transf_u_inv_90}
\end{align}
with
\begin{align}
 \sin^2\theta &= \cos^2\theta' \cos^2\varphi' + \sin^2 \varphi', \\
 \sin^2\theta' &= \cos^2 \theta \cos^2 \varphi + \sin^2 \varphi.
\end{align}
Then, the amplitudes $U^{e/o}_n$ and $V^{e/o}_n$ of the acoustic and shear modes can be obtained via
\begin{align}
 U^{e/o}_n &= \int_0^{2\pi} \frac{d\varphi'}{2\pi} \int_0^\pi \frac{d\theta' \sin\theta'}{\sin^2\theta} \nonumber\\
 & \times 
 (-\cos\theta' \cos\varphi' u'^{\hat{\theta}'} + \sin\varphi' u'^{\hat{\varphi}'}) F^{e/o}_n(\theta), \label{eq:transf_U_90}\\
 V^{e/o}_n &= \int_0^{2\pi} \frac{d\varphi'}{2\pi} \int_0^\pi d\theta' \sin\theta' \nonumber\\
 & \times 
 (-\sin\theta' u'^{\theta'} - \cos\theta' \cos\varphi' u'^{\hat{\varphi}'}) G^{e/o}_n(\theta).
 \label{eq:transf_V_90}
\end{align}
For completeness, in Fig.~\ref{fig:isotropy_test} we report the results obtained repeating the benchmark in Sec.\ref{sec:numerics:sound} on the rotated grid. The figure highlights in a quantitative way that the results on the rotated grid are consistent with those presented in the main text, and in particular, the fifth-order convergence of the \ac{weno5} advection scheme is preserved.

\begin{figure}[tb]
 \includegraphics[width=0.99\columnwidth]{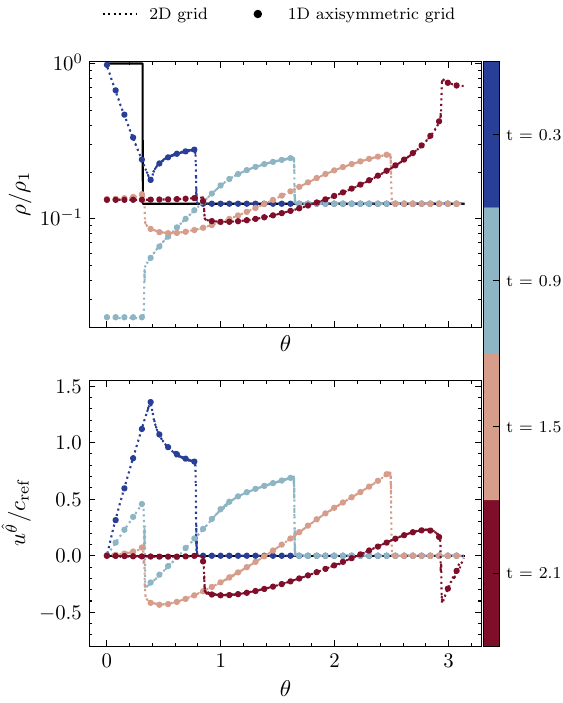}
    \caption{Riemann problem: same as Fig.~\ref{fig:riemann}, with the points showing the 1D (axisymmetric) solution discussed in Sec.~\ref{sec:numerics:riemann} and the dotted lines showing the $2D$ solution corresponding to the rotated grid introduced in Eq.~\eqref{eq:transf_90}.
    }
    \label{fig:riemann_2D}
\end{figure}

In Fig.~\ref{fig:riemann_2D}, we further consider a test of the isotropy of our scheme by simulating the 
shock wave problem considered in Sec.~\ref{sec:numerics:riemann} on a grid rotated according to Eq.~\eqref{eq:transf_90}. 
To ensure the same resolution, we considered a grid with $N_\theta \times N_\varphi$, with $(N_\theta, N_\varphi) = (1024, 1)$ 
points in the 1D (axisymmetric) case and $N_\theta = N_\varphi = 1024$ for the 2D (rotated grid) simulation. 
The excellent agreement between the two simulations confirms the isotropy of our scheme.

\section{Euler solver}\label{app:Euler}

In this section of the appendix, we discuss the Flux Vector Splitting method for the numerical solution of the 
Euler Eqs. in the axisymmetric case, obtained by setting $\eta = \zeta = 0$ in Eqs.~\eqref{eq:axi_continuity}--\eqref{eq:axi_NSE_uth}: 
\begin{align}
    \frac{\partial \rho}{\partial t} + \frac{\partial_{\theta}(\rho u^{\hat{\theta}} \sin\theta)}{R \sin\theta} &= 0, \nonumber\\
    \frac{\partial(\rho u^{\hat{\theta}})}{\partial t} + \frac{\partial_{\theta}[\rho (u^2_{\hat{\theta}} + c_s^2) \sin\theta]}{R \sin\theta} &= \frac{\rho c_s^2}{R \tan\theta},
    \label{eq:Euler_cons}
\end{align}
where we took into account that the pressure is $p = c_s^2 \rho$ for an isothermal gas.

Considering the vector $U = (\rho, \rho u^{\hat{\theta}})^T$ of primary variables, we consider the following abstract form of the Euler equations:
\begin{equation}
 \partial_t \mathbf{U} + \frac{\partial_\theta (\mathbf{F} \sin \theta)}{R \sin\theta} = \mathbf{S}_{\rm inv},
\end{equation}
where the flux vector and the inviscid source vector read
\begin{equation}
 \mathbf{F} = \begin{pmatrix}
  \rho u^{\hat{\theta}} \\ 
  \rho(u_{\hat{\theta}}^2 + c_s^2)
 \end{pmatrix}, \quad 
  \mathbf{S}_{\rm inv} = \begin{pmatrix}
 0 \\ 
 (\rho c_s^2 / R) \cot\theta
  \end{pmatrix}.
\end{equation}
The Jacobian matrix $\mathbb{A} = \partial \mathbf{F} / \partial \mathbf{U}$ reads 
\begin{equation}
 \mathbb{A} 
 = \begin{pmatrix}
     0 & 1 \\ 
      c_s^2 - u_{\hat{\theta}}^2 & 2u^{\hat{\theta}}
 \end{pmatrix},
\end{equation}
having the eigenvalues $\lambda_\pm = u^{\hat{\theta}} \pm c_s$. We take the corresponding right- and left-eigenvectors as
\begin{equation}
 \mathbf{R}_\pm = \begin{pmatrix}
  1 \\ u^{\hat{\theta}} \pm c_s
 \end{pmatrix}, \quad 
 \mathbf{L}_\pm = \frac{1}{2c_s} \begin{pmatrix}
     c_s \mp u^{\hat{\theta}} \\ \pm 1
 \end{pmatrix},
\end{equation}
satisfying $\mathbb{A} \mathbf{R}_\pm = \lambda_\pm \mathbf{R}_\pm$ and $\mathbf{L}_{\pm} \mathbb{A} = \lambda_\pm \mathbf{L}^T_\pm$. We then introduce the matrices $\mathbb{R}$ and $\mathbb{L}$ of right- and left-eigenvectors via
\begin{subequations}
\begin{align}
 \mathbb{R} &= 
 \begin{pmatrix}
 \mathbf{R}_+ & \mathbf{R}_-\end{pmatrix} = 
 \begin{pmatrix}
  1 & 1 \\ 
  u^{\hat{\theta}} + c_s & u^{\hat{\theta}} - c_s 
 \end{pmatrix}, \\
 \mathbb{L} &= 
 \begin{pmatrix}
 \mathbf{L}^T_+ \\\medskip \mathbf{L}^T_-\end{pmatrix} = \frac{1}{2c_s}
 \begin{pmatrix}
  c_s - u^{\hat{\theta}} & 1 \\
  c_s + u^{\hat{\theta}} & -1
 \end{pmatrix},
\end{align}
\end{subequations}
satisfying $\mathbb{A} \mathbb{R} = \mathbb{R} \boldsymbol{\Lambda}$ and $\mathbb{L} \mathbb{A} = \boldsymbol{\Lambda} \mathbb{L}$, with $\boldsymbol{\Lambda} = {\rm diag}(\lambda_+, \lambda_-)$ being the diagonal form of the Jacobian matrix $\mathbb{A}$.

We solve the Euler equations using the third-order Runge-Kutta time-stepping method introduced in Sec.~\ref{sec:method:time}. The advection is computed at each point $s$ ($1 \le s \le N_\theta$), having polar coordinate $\theta_s$, using a Komissarov-type scheme:
\begin{equation}
 \left[\frac{\partial_\theta (\mathbf{F} \sin\theta)}{R \sin\theta}\right]_s = \frac{\mathbf{F}_{s + \frac{1}{2}} \sin \theta_{s +\frac{1}{2}} - \mathbf{F}_{s-\frac{1}{2}} \sin\theta_{s-\frac{1}{2}}}{R \delta \theta \sin\theta_s},
\end{equation}
with $\mathbf{F}_{s \pm 1/2}$ being the fluxes at the two interfaces $s \pm 1/2$ of cell $s$. 
For brevity, we focus on the flux through the interface at $s + 1/2$. Here, we require the Jacobian matrix $\mathbb{A}_{s + 1/2}$, computed for the primitive variables at the cell interface, defined as $\mathbf{U}_{s + 1/2} = \frac{1}{2} (\mathbf{U}_s + \mathbf{U}_{s + 1})$. Then, the flux $\mathbf{F}_{s + 1/2}$ is reconstructed using the matrix of right eigenvectors $\mathbb{R}_{s + 1/2}$,
\begin{equation}
 \mathbf{F}_{s + 1/2} = \mathbb{R}_{s + 1/2} \mathbf{Q}_{s + 1/2},
\end{equation}
with the vector of characteristic fluxes $\mathbf{Q}_{s + 1/2} \equiv (Q^+_{s + 1/2}, Q^-_{s + 1/2})^T$ computed in an upwind-biased manner using the WENO-5 advection scheme, described in Eqs.~\eqref{eq:fluxes_W5}--\eqref{eq:WENO5_sigma}. The values $\mathbf{Q}_{s'}$  ($s - 3 \le s' \le s + 3$) required for the WENO-5 stencils in Eqs.~\eqref{eq:fluxes_W5} are computed using the matrix of left eigenvectors, as follows: $\mathbf{Q}_{s'} = \mathbb{L}_{s + 1/2} \mathbf{F}_{s'}$. The upwind direction is given for each flux $Q^\pm_{s + 1/2}$ by the corresponding eigenvalue, $\lambda^\pm_{s + 1/2}$. For the ghost nodes outside the grid, we employ $(\rho_{1 - s'}, u^{\hat{\theta}}_{1-s'}) = (\rho_{s'}, -u^{\hat{\theta}}_{s'})$ and $(\rho_{N_\theta + s'}, u^{\hat{\theta}}_{N_\theta + s'}) = (\rho_{N_\theta + 1 - s'}, -u^{\hat{\theta}}_{N_\theta + 1 - s'})$, with $1 \le s' \le 3$.

\bibliography{biblio}

\end{document}